\documentclass[journal]{IEEEtran}
\usepackage{amsmath}
\usepackage{amssymb}
\usepackage{makecell}

\usepackage{graphicx}
\usepackage{epstopdf}
\usepackage{amsmath}
\usepackage{array}
\newcommand{\PreserveBackslash}[1]{\let\temp=\\#1\let\\=\temp}
\newcolumntype{C}[1]{>{\PreserveBackslash\centering}p{#1}}
\newcolumntype{R}[1]{>{\PreserveBackslash\raggedleft}p{#1}}
\newcolumntype{L}[1]{>{\PreserveBackslash\raggedright}p{#1}}
\usepackage[usenames]{color}
\usepackage{colortbl,booktabs}
\usepackage{tabularx}
\usepackage{multicol}
\usepackage{booktabs}
\usepackage[Symbol]{upgreek}
\usepackage{bm}
\usepackage{cite}
\usepackage{balance}
\usepackage{mathrsfs}
\usepackage{url}
\usepackage{algorithm} %format of the algorithm
\usepackage{algorithmic} %format of the algorithm
\usepackage{threeparttable}
\usepackage{amsmath}
\usepackage{dcolumn}
\usepackage{multirow}
\usepackage{booktabs}
\usepackage{amsfonts}
\newcommand{\tabincell}[2]{\begin{tabular}{@{}#1@{}}#2\end{tabular}}
\begin{document}
	\title{Beamspace Precoding and Beam Selection for Wideband Millimeter-Wave MIMO Relying on Lens Antenna Arrays \vspace{-1mm}}
	%
	%\author{\IEEEauthorblockN{Wenqian Shen, Linglong Dai, and Zhaocheng Wang}\vspace{-0mm}}
	\author{Wenqian Shen,
		Xiangyuan Bu, Xinyu Gao, Chengwen Xing, and Lajos Hanzo,~\IEEEmembership{Fellow,~IEEE}
		\thanks{W. Shen, X. Bu, and C. Xing are with School of Information and Electronics, Beijing Institute of Technology, Beijing 100081, China (e-mails: \{wshen, bxy, chengwenxing\}@bit.edu.cn).}
		\thanks{X. Gao is with the Department of Electronic Engineering, Tsinghua University, Beijing 100084, China (e-mail: xy-gao14@mails.tsinghua.edu.cn).}
		\thanks{L. Hanzo is with the Department of Electronics and Computer Science, University of Southampton, Southampton SO17 1BJ, UK (e-mail: lh@ecs.soton.ac.uk).}
		%\thanks{A. Sayeed is with the Department of Electrical and Computer Engineering, University of Wisconsin, Madison, WI 53706, USA (email: akbar@engr.wisc.edu).}
%		\thanks{This work was supported by the Royal Academy of Engineering under the UK-China Industry Academia Partnership Programme Scheme (Grant No. UK-CIAPP${\backslash}$49). L. Hanzo would also like to acknowledge the financial support of the ERC Advanced Fellow Grant.} \vspace{-0mm}
%       \thanks{This work was supported by ?}
\vspace{-3mm}	
}
	\maketitle
\begin{abstract}
Millimeter-wave (mmWave) multiple-input multiple-out (MIMO) systems relying on lens antenna arrays are capable of achieving a high antenna-gain at a considerably reduced number of radio frequency (RF) chains via beam selection. However, the traditional beam selection network suffers from significant performance loss in wideband systems due to the effect of beam squint. In this paper, we propose a phase shifter-aided beam selection network, which enables a single RF chain to support multiple focused-energy beams, for mitigating the beam squint in wideband mmWave MIMO systems. Based on this architecture, we additionally design an efficient transmit precoder (TPC) for maximizing the achievable sum-rate, which is composed of beam selection and beamspace precoding. Specifically, we decouple the design problems of beamspace precoding and beam selection by exploiting the fact that the beam selection matrix has a limited number of candidates. For the beamspace precoding design, we propose a successive interference cancellation (SIC)-based method, which decomposes the associated optimization problem into a series of subproblems and solves them successively. For the beam selection design, we propose an energy-max beam selection method for avoiding the high complexity of exhaustive search, and derive the number of required beams for striking an attractive trade-off between the hardware cost and system performance. Our simulation results show that the proposed beamspace precoding and beam selection methods achieve both a higher sum-rate and a higher energy efficiency than its conventional counterparts.
	\end{abstract}
	
	\IEEEpeerreviewmaketitle
	
\section{Introduction}\label{S1}
\IEEEPARstart{M}{illimeter}-wave (mmWave) communication has become a key technique for next-generation wireless communication systems owing to its substantial bandwidth\cite{alkhateeb2014mimo,CST_SABusari_Millimeter,CM_YNiu_CrosslayerMmWave}, but unfortunately it has a high free-space path loss \cite{Access_TSRappaport_mmwave}. A promising technique for mitigating this problem is to involve massive multiple-input-multiple-output (MIMO) techniques \cite{JTSP_HRobert_OverviewMmwave,JSAC_XZhai_joint,CM_JZhang_onlow}. Fortunately the mm-scale wavelengths allow for example 256 antenna elements to be packed in a relatively small physical size \cite{Access_TSRappaport_mmwave}, which is capable of compensating for the high path loss with the aid of transmit precoder (TPC) \cite{JTSP_HRobert_OverviewMmwave}. Despite this potential, a range of practical challenges hamper the implementation of mmWave massive MIMO systems. Traditionally, MIMO systems tend to rely on the fully digital precoding, where each antenna is supported by a dedicated radio frequency (RF) chain. This leads to high hardware cost and high power consumption for mmWave massive MIMO systems relying on large antenna arrays\cite{TSP_ALiu_phase}.

Hence, hybrid TPC solutions have been proposed for circumventing this problem \cite{el2013spatially,TWC_AAhmed_LimitedHybridPrecoding,JSAC_KVenugopal_Wideband,JSAC_XGao_EnergyEfficient,JSTSP_FSohrabi_hybrid16}, where the TPC is decomposed into the digital precoding requiring a reduced number of RF chains and the large analog precoding, realized by analog phase shifters. Hybrid precoding combined with beam selection has been discussed in \cite{TSP_VVRatnam_hybrid} and \cite{TVT_Spayami_hybrid}, when the phase shifters are used for analog precoding. A switching network is incorporated between the RF chains and phase shifters in \cite{TSP_VVRatnam_hybrid} for reducing the hardware cost, while retaining good performance. The authors of \cite{TVT_Spayami_hybrid} propose to add a fully-/sub- connected switching network between the phase shifters and the antennas for reducing the number of phase shifters without degrading the spectral efficiency.
A particularly promising way of implementing the analog precoding is using the lens antenna array
\cite{TAP_JBrady_Beamspace,TCOM_PAmardori_BeamSelection,JSAC_YZeng_Electromagnetic,TCOM_YZeng_PDMA,JSTSP_RGuo_Joint}, which includes a lens and an antenna array located on the focal surface of the lens.
Lenses focus the incident mmWave beams (signals) on different antennas.
In this way, the traditional spatial channel is transformed into the so-called ``beamspace channel." Due to the limited scattering experienced at mmWave frequencies, the number of focused-energy beams of the beamspace channel is small \cite{TCOM_TSRappaport_wideband}.
Thus the transmitter can select a subset of focused-energy beams by switches for ensuring that the number of RF chains is reduced without significant performance loss\cite{CL_XGao_near,CL_CFeng_beam}. Then, the digital precoding is performed on the reduced-dimensional beamspace channel, which is termed as ``beamspace precoding'' in this paper.

However, designing the TPC consisting of beam selection and beamspace precoding is not a trivial task for wideband mmWave MIMO systems. Due to the effect of beam squint in wideband systems \cite{ICC_XGao_beamspace}, the focused-energy beams of beamspace channels become frequency-dependent. However, the traditional beam selection network is frequency-independent \cite{JSAC_KVenugopal_Wideband}, which will lead to considerable performance loss due to the power leakage of beamspace channel at certain frequencies. To overcome this problem, an intuitive technique is to select more beams to cover the entire channel bandwidth. However, this will unfortunately increase the number of RF chains, hence the hardware cost and power consumption as well. In conclusion, it is challenging to design the beam selection and beamspace precoding schemes for wideband systems to maintain satisfactory performance across the entire channel bandwidth without increasing the number of RF chains.

\subsection{Prior work}
For the family of wideband mmWave MIMO systems relying on lens antenna arrays, the authors of \cite{TWC_WHuang_Wideband,TWC_YZeng_Multiuser} adopted a single-carrier transmission scheme relying on a path delay-compensation technique, where the frequency-selective multi-user MIMO channels are transformed to several low-dimensional parallel frequency-flat MIMO channels for different users. Then, a joint antenna selection and beamforming scheme was proposed for eliminating the inter-user interference.
%However, the effect of beam squint experienced in wideband mmWave MIMO systems was not taken into consideration.
The authors of \cite{ICCW_JHBrady_wideband} analyzed the effect of beam squint on the system's performance loss and quantified the number of dominant beams required for maintaining a satisfactory performance. The results were obtained for single-input multiple-output systems communicating over line-of-sight channels. Gao et al. \cite{TWC_XGao_BeamspaceChannelEstimation} proposed an adaptive beam selection network, which consists of a small number of 1-bit phase shifters. This architecture is used adaptively as a random combiner during the channel estimation while during data transmission as a traditional beam selection network. However, the performance erosion caused by the effect of beam squint was still not alleviated.
%
%For the traditional hybrid TPC systems based on the phase shifter array, some wideband TPC methods were proposed \cite{TCOM_AAlkhateeb_Frequency,JSAC_FSohrabi_hybrid17,TWC_SPark_Dynamic,TSP_Ylin_hybrid}. In these methods, the analog TPC matrix is frequency-independent while the digital precoding matrix is frequency-dependent. \cite{TCOM_AAlkhateeb_Frequency} proposed a hybrid TPC design given the limited channel feedback between the transmitter and receiver.
%\cite{JSAC_FSohrabi_hybrid17} proposed a unified TPC design for both the fullyy connected and the sub-array connected hybrid TPC structures to maximize the overall throughput.
%\cite{TWC_SPark_Dynamic} proposed a dynamic sub-array connected hybrid TPC structure that adapts the subarray structure according to the long-term channel statistics.
%\cite{TSP_Ylin_hybrid} designed the subcarrier beamformers based on spectral analysis of the covariance matrices of the subcarrier channels.
%These methods are not directly applicable to the beamspace precoding in wideband mmWave MIMO systems with the lens antenna array due to the different hardware constrains between the lens antenna array and the phase shifter array.

\subsection{Contributions}
In this paper, we propose a phase shifter-aided selection network combined with an efficient TPC design for wideband mmWave MIMO systems relying on lens antenna arrays. The main contributions of this paper are summarized as follows:
\begin{itemize}
	\item We propose a phase shifter-aided selection network for mitigating the effect of beam squint. The wideband beamspace MIMO channel is frequency-dependent due to the beam squint, while the beam selection network is frequency-independent, which will lead to the power leakage of beamspace channel at certain frequencies. In the proposed selection network, we capture most of the channel's output energy over the entire bandwidth without increasing the number of RF chains, where each RF chain is designed for supporting multiple focused-energy beams via a sub-array connected phase shifter network. Upon relying on a carefully designed TPC composed of beamspace precoding and beam selection, the proposed architecture achieves a near-unimpaired sum-rate despite relying on a reduced number of RF chains.
	
	\item To design an efficient TPC maximizing the sum-rate, we first decouple the design problems of beamspace precoding and beam selection by exploiting the fact that the number of candidate beam selection matrices is limited. For a given beam selection matrix, we propose a successive interference cancellation (SIC)-based beamspace precoding scheme, which is capable of achieving the maximum mutual information (MI) of wideband mmWave MIMO channels. Specifically, the beamspace precoding is realized by intrinsically amalgamating our baseband precoding and the phase shifter network. Given the sub-array connected structure of the phase shifter network, the optimization problem of MI maximization may be readily decomposed into several subproblems. Then, by appropriately adapting the classic concept of SIC signal detection \cite{JSAC_XGao_EnergyEfficient}, we propose a SIC-based beamspace precoding design, where each subproblem is solved after removing the contributions of the previously solved subproblems.
	
	\item Once the beamspace precoding method has been designed for a given beam selection matrix, the TPC design problem is reduced to the beam selection design problem, which can be solved by the optimal exhaustive search, but its complexity may still be excessive. To avoid the high complexity of exhaustive search, we develop an energy-max beam selection method. Specifically, we first prove that the energy-max beam selection design is capable of approaching the maximum MI, which means that the selection matrix should be designed to select the focused-energy beams of the wideband beamspace channel. Then, we derive the number of beams required for striking an attractive hardware cost/power consumption vs sum-rate trade-off. Extensive simulations verify the superior performance of the proposed beamspace precoding and beam selection methods.
\end{itemize}

The rest of the paper is organized as follows. In Section II, we present both our system and channel models. In Section III, we first propose a phase shifter-aided selection network for mmWave MIMO systems relying on lens antenna arrays. Then the SIC-based beamspace precoding and energy-max beam selection methods are developed. Our simulation results are provided in Section IV. Finally, our conclusions are drawn in Section V.

\emph{Notation}: Lower-case and upper-case boldface letters denote
vectors and matrices, respectively.  $(\cdot)^{\rm{T}}$,
$(\cdot)^{\rm H}$, $(\cdot)^{\rm *}$, and $(\cdot)^{-1}$ denote the transpose,
conjugate transpose, conjugate, and inverse of a matrix, respectively.
$\left|\cdot \right|$ denotes the determinant of a matrix. $\lceil \cdot \rceil$ denotes rounding toward its nearest higher integer.
${\rm tr}(\cdot)$ is the trace of a matrix.
%$\mathbf{\Phi}^{\dagger}=\mathbf{\Phi}(\mathbf{\Phi}^{\rm H}\mathbf{\Phi})^{-1}$ is the Moore-Penrose pseudo-inverse.
$\left\| \cdot\right\|_{\rm F}$ is the Frobenius norm of a matrix. Finally, $\mathbf{I}_N$ denotes the identity matrix of size $N\times N$.

\section{System and Channel Models}\label{S2}
In this section, we describe our wideband mmWave MIMO system relying on a lens antenna array and wideband beamspace channel model. The effect of beam squint over different frequencies is also highlighted.
\subsection{System model}
We commence by briefly introducing the wideband mmWave MIMO system model relying on a lens antenna array.
We consider a MIMO aided orthogonal-frequency-division-multiplexing (OFDM) system using $K$ subcarriers.
There are $N_{\rm t}$ transmit antennas (TAs) and $N_{\rm t}^{\rm RF}$ RF chains at the transmitter ($N_{\rm t}\ge N_{\rm t}^{\rm RF}$). At the receiver side, there are $N_{\rm r}$ receive antennas (RAs) and $N_{\rm r}^{\rm RF}$ RF chains ($N_{\rm r} \ge N_{\rm r}^{\rm RF}$). The transmitter conveys   $N_{\rm s}$ data streams at each subcarrier to the receiver, so that we have $N_{\rm s}\le N_{\rm t}^{\rm RF}$ and $N_{\rm s}\le N_{\rm r}^{\rm RF}$.

\begin{figure}[t]
	\center{\includegraphics[width=1\columnwidth] {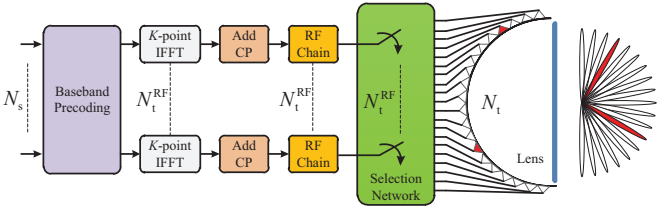}}
	\caption{Illustration of the transmitter in the wideband mmWave MIMO system relying on a lens antenna array. After baseband precoding, the precoded data streams are transformed into the time domain using the IFFT. After adding CP before each OFDM symbol, the time-domain signals are transmitted through a subset of antennas selected by a selection network.}
	\label{Fig_OFDM_Beamspace_MIMO}
\end{figure}
The transmitter is illustrated in Fig. \ref{Fig_OFDM_Beamspace_MIMO}, while the receiver relies on the inverse architecture. For the $k$-th subcarrier, the transmit data $\mathbf{s}[k]\in\mathbb{C}^{N_{\rm s}\times 1}$ ($k=1,2,\cdots, K$) is first precoded for reducing the interference between the different data streams. The baseband precoding matrix $\mathbf{F}_{\rm BB}[k]\in\mathbb{C}^{N_{\rm t}^{\rm RF}\times N_{\rm s}}$ can adjust both the amplitude and phase of each data stream. Then the precoded data $\mathbf{F}_{\rm BB}[k]\mathbf{s}[k]$ is transformed into the time domain using the $K$-point inverse fast Fourier transform (IFFT). Then the cyclic prefix (CP) is added to each data block to eliminate the inter-symbol interference (ISI). A selection network $\mathbf{S}_{\rm t}\in\mathbb{C}^{N_{\rm t}\times N_{\rm t}^{\rm RF}}$ selects $N_{\rm t}^{\rm RF}$ TAs to be coupled to $N_{\rm t}^{\rm RF}$ RF chains through mmWave switches\footnote{Note that the switches are passive devices, which will lead to inevitable insertion loss compared to active phase shifters. Fortunately, mmWave switches exhibit low insertion loss ($\approx$1dB) and good isolation properties \cite{Access_RMendez_hybridmimo}.}.
It is worth noting that the selection network $\mathbf{S}_{\rm t}$ is frequency-independent. The transmitted discrete-time complex baseband signal $\mathbf{x}[k]\in\mathbb{C}^{N_{\rm t}\times 1}$ at the $k$-th subcarrier is given by \cite{WCL_WShen_Channelfeedback}
\begin{align}\label{eq_xk}
\mathbf{x}[k]=\sqrt{\rho/N_{\rm s}}\mathbf{S}_{\rm t}\mathbf{F}_{\rm BB}[k]\mathbf{s}[k],
\end{align}
where $\rho$ is the transmit power. The transmit data $\mathbf{s}[k]$ has the normalized power of ${\rm E}\left[ \mathbf{s}[k]\mathbf{s}^{\rm H}[k]\right] =\mathbf{I}_{N_{\rm s}}$. The baseband precoding matrix satisfies the transmit power constraint of ${\rm tr}\left(\mathbf{F}^{\rm H}_{\rm BB}[k]\mathbf{F}_{\rm BB}[k] \right)=N_{\rm s}$. The selection matrix $\mathbf{S}_{\rm t}$ has one and only one non-zero element ``1" in each column, so that the RF signal generated by a single RF chain is transmitted by a selected antenna. The transmit signals are passed through a lens to form several focused-energy beams. The lens transforms the spatial channel into the beamspace channel. We will discuss the lens and the beamspace channel later in the following Subsection II-B.

At the receiver, the incident beams (signals) are focused on a subset of the RAs by a lens. Those antennas associated with focused-energy beams are selected by a selection network $\mathbf{S}_{\rm r}\in\mathbb{C}^{N_{\rm r}\times N_{\rm r}^{\rm RF}}$ to be supported by $N_{\rm r}^{\rm RF}$ RF chains. After passing through the RF chains, the received baseband signals are transformed back to the frequency domain using the $K$-point FFT. Then, the symbols at each subcarrier are combined by a baseband combining matrix $\mathbf{W}_{\rm BB}[k]\in\mathbb{C}^{N_{\rm r}^{\rm RF}\times N_{\rm s}}$. The constraints imposed on the selection matrix $\mathbf{S}_{\rm r}$ and combining matrix $\mathbf{W}_{\rm BB}[k]$ at the receiver are similar to those of the transmitter. Assuming that the frame synchronization/timing offset (TO) synchronization and carrier frequency offset (CFO) synchronization are perfect at the receiver, the received discrete-time complex baseband signal $\mathbf{y}[k]\in\mathbb{C}^{N_{\rm s}\times 1}$ at the $k$-th subcarrier is then given by
%\begin{align}\label{eq_yk}
%	\mathbf{y}[k]=&\sqrt{\rho/N_{\rm s}}\mathbf{W}^{\rm H}_{\rm BB}[k]\mathbf{S}^{\rm H}_{\rm r}\mathbf{H}_{\rm b}[k]\mathbf{S}_{\rm t}\mathbf{F}_{\rm BB}[k]\mathbf{s}[k]\nonumber\\&+
%	\mathbf{W}^{\rm H}_{\rm BB}[k]\mathbf{S}^{\rm H}_{\rm r}\mathbf{n}[k],
%\end{align}
\begin{align}\label{eq_yk}
\mathbf{y}[k]=\mathbf{W}^{\rm H}_{\rm BB}[k]\mathbf{S}^{\rm H}_{\rm r}\mathbf{H}_{\rm b}[k]\mathbf{x}[k]+
\mathbf{W}^{\rm H}_{\rm BB}[k]\mathbf{S}^{\rm H}_{\rm r}\mathbf{n}[k],
\end{align}
where $\mathbf{H}_{\rm b}[k]\in\mathbb{C}^{N_{\rm r}\times N_{\rm t}}$ is the beamspace channel at the $k$-th subcarrier generated by the lens, which will be discussed in the following Subsection II-B. Finally, $\mathbf{n}[k]\sim \mathcal{N}(0,\sigma^2\mathbf{I}_{N_{\rm r}})$ is the additive white Gaussian noise (AWGN), where $\sigma^2$ is the noise power.

\subsection{Beamspace channel model}
To incorporate the multi-path structure of wideband mmWave MIMO channels, we adopt a ray-based channel model having $L$ clusters of scatterers\cite{CambridgeUP_TDavid_Fundamentals,TCOM_AAlkhateeb_Frequency}. Each cluster has a limited angle-of-departure/arrival (AoD/AoA) spread $\psi_{\rm t}^\ell$ and $\psi_{\rm r}^\ell$. The $\ell$-th cluster is assumed to be contributed by $S_{\ell}$ propagation subpaths. Each subpath has a time delay $\tau_{s_\ell}$, a physical AoD $\theta_{\rm t}^{s_{\ell}} \in \psi_{\rm t}^\ell$, a physical AoA $\theta_{\rm r}^{s_{\ell}}\in \psi_{\rm r}^\ell$, and a complex path gain $\alpha_{s_{\ell}}$ ($s_{\ell}=1,2,\cdots, S_{\ell}$). The spatial AoD and AoA at the $k$-th subcarrier are defined as $\phi_{\rm t}^{s_{\ell},k} =\left(d/\lambda_k\right)\sin\theta_{\rm t}^{s_{\ell}}$ and $\phi_{\rm r}^{s_{\ell},k}=\left(d/\lambda_k\right)\sin\theta_{\rm r}^{s_{\ell}}$, where $d$ is the antenna spacing and $\lambda_k$ is the wavelength of the $k$-th subcarrier.
Under this model, the delay-$d$ tap $\mathbf{H}_d\in\mathbb{C}^{N_{\rm r}\times N_{\rm t}}$ of the wideband mmWave MIMO channel can be expressed as \cite{TCOM_AAlkhateeb_Frequency,JSAC_KVenugopal_Wideband}
\begin{align}\label{eq_Hd} % eq1
\mathbf{H}_d=\sum_{\ell=1}^{L}\sum_{s_{\ell}=1}^{S_{\ell}}\alpha_{s_{\ell}}p_{\rm rc}(dT_{\rm s}-\tau_{s_\ell})
\mathbf{a}_{\rm r}\left( {\phi_{\rm r}^{s_{\ell},k}} \right)  \mathbf{a}_{\rm t}^{\rm H} \left( {\phi_{\rm t}^{s_{\ell},k}} \right),
\end{align}
where $p_{\rm rc}(\tau)$ is a band-limited pulse-shaping filter evaluated at $\tau$, with the system's sampling period given by $T_{\rm s}$. Furthermore, $\mathbf{a}_{\rm t}\left( \phi_{\rm t}^{s_{\ell},k}\right) \in\mathbb{C}^{N_{\rm t} \times 1}$ and $\mathbf{a}_{\rm r}\left( \phi_{\rm r}^{s_{\ell},k}\right) \in\mathbb{C}^{N_{\rm r} \times 1}$ are the antenna array responses at the transmitter and receiver, which can be respectively presented as follows, where the uniform linear arrays (ULAs) are employed
\begin{align}
&\mathbf{a}_{\rm t}\left( \phi_{\rm t}^{s_{\ell},k}\right) =\frac{1}{\sqrt{N_{\rm t}}}\left[ 1,e^{j2\pi\phi_{\rm t}^{s_{\ell},k}},\cdots, e^{j2\pi(N_{\rm t}-1)\phi_{\rm t}^{s_{\ell},k}}\right] ^{\rm H},\\
&\mathbf{a}_{\rm r}\left( \phi_{\rm r}^{s_{\ell},k}\right) =\frac{1}{\sqrt{N_{\rm r}}}\left[ 1,e^{j2\pi\phi_{\rm r}^{s_{\ell},k}},\cdots, e^{j2\pi(N_{\rm r}-1)\phi_{\rm r}^{s_{\ell},k}}\right] ^{\rm H}.
\end{align}

Given the $\mathbf{H}_d$ in (\ref{eq_Hd}), the spatial wideband mmWave MIMO channel $\mathbf{H}[k]\in\mathbb{C}^{N_{\rm r}\times N_{\rm t}}$ at the $k$-th subcarrier is described as
%\begin{align}\label{eq_Hk} % eq1
%	\mathbf{H}[k]&=\sum_{d=0}^{D-1}\mathbf{H}_d e^{-j2\pi dk/K}\\\nonumber
%	&=\sum_{\ell=1}^{L}\sum_{s_{\ell}=1}^{S_{\ell}}\beta_{s_{\ell,k}}
%	\mathbf{a}_{\rm r}(\phi_{\rm r}^{s_{\ell}})\mathbf{a}_{\rm t}^{\rm H}(\phi_{\rm t}^{s_{\ell}}),
%\end{align}
\begin{align}\label{eq_Hk} % eq1
\mathbf{H}[k]&=\sum_{d=0}^{D-1}\mathbf{H}_d e^{-j2\pi dk/K}\\\nonumber
&=\sum_{\ell=1}^{L}\sum_{s_{\ell}=1}^{S_{\ell}}\beta_{s_{\ell,k}}
\mathbf{a}_{\rm r}(\phi_{\rm r}^{s_{\ell},k})\mathbf{a}_{\rm t}^{\rm H}(\phi_{\rm t}^{s_{\ell},k}),
\end{align}
where $\beta_{s_{\ell,k}}=\sum_{d=0}^{D-1}\alpha_{s_{\ell}}p_{\rm rc}(dT_{\rm s}-\tau_{s_\ell}) e^{-j2\pi dk/K}$ is the complex gain at the $k$-th subcarrier.
In contrast to narrowband systems, the spatial AoD/AoA ($\phi_{\rm t}^{s_{\ell},k}$/$\phi_{\rm r}^{s_{\ell},k}$) varies with the subcarrier index $k$ in wideband systems. Since we have $\lambda_k=c/f_k$, where $c$ is the speed of light and $f_k$ is the frequency of the $k$-th subcarrier, the spatial AoD can be rewritten as $\phi_{\rm t}^{s_{\ell},k}=\left(df_k/c\right)\sin\theta_{\rm t}^{s_{\ell}}$
and the spatial AoA can be rewritten as $\phi_{\rm r}^{s_{\ell},k}=\left(df_k/c\right)\sin\theta_{\rm r}^{s_{\ell}}$. The value $f_k$ is given by $f_k = {f_{\rm c}} + \frac{B}{K}\left( {k - 1 - \frac{{K - 1}}{2}} \right)$ with $f_{\rm c}$ and $B$ being the central carrier frequency and the system's bandwidth, respectively.

Due to the frequency-dependent spatial AoD/AoA, the beamspace channel of wideband systems is different from that in narrowband systems. Specifically, by exploiting the fact that a lens computes a spatial Fourier transformation of incident signals, the spatial channel $\mathbf{H}[k]$ is transformed into its beamspace representation $\mathbf{H}_{\rm b}[k]$ as \cite{WCL_WShen_Channelfeedback,TWC_XGao_BeamspaceChannelEstimation}
\begin{align}\label{eq_Hbk}
\mathbf{H}_{\rm b}[k]=\mathbf{U}_{\rm r}^{\rm H}\mathbf{H}[k]\mathbf{U}_{\rm t},
\end{align}
where ${\bf{U}}_{\rm{t}} = \left[ {\bf{a}}_{\rm{t}}\left(\bar\phi_{\rm{t}}^1 \right),{\bf{a}}_{\rm{t}}\left(\bar\phi_{\rm{t}}^2\right), \cdots,{\bf{a}}_{\rm{t}}\left(\bar\phi _{\rm{t}}^{N_{\rm{t}}} \right) \right]$ and ${\bf{U}}_{\rm{r}} = \left[ {\bf{a}}_{\rm{r}}\left(\bar\phi_{\rm{r}}^1 \right),{\bf{a}}_{\rm{r}}\left(\bar\phi_{\rm{r}}^2\right), \cdots,{\bf{a}}_{\rm{r}}\left(\bar\phi _{\rm{r}}^{N_{\rm{r}}} \right) \right]$ are discrete Fourier transform (DFT) matrices, and $\bar\phi_{\rm{t}}^n = \frac{1}{N_{\rm{t}}}\left( n - \frac{N_{\rm{t}} + 1}{2} \right)$ for $n = 1,2,\cdots,N_{\rm{t}}$ and $\bar \phi_{\rm{r}}^n = \frac{1}{N_{\rm{r}}} \left( n - \frac{{N_{\rm{r}}} + 1}{2} \right)$ for $n = 1,2, \cdots,{N_{\rm{r}}}$.
By substituting (\ref{eq_Hk}) into (\ref{eq_Hbk}), we have
\begin{align}\label{eq_Hbk2}
\mathbf{H}_{\rm b}[k]&=\sum_{\ell=1}^{L}\sum_{s_{\ell}=1}^{S_{\ell}}\beta_{s_{\ell,k}}
\mathbf{U}_{\rm r}^{\rm H}
\mathbf{a}_{\rm r}\left( \phi_{\rm r}^{s_{\ell},k}\right)  {\bf{a}}_{\rm{t}}^{\rm{H}}\left( {\phi _{\rm{t}}^{{s_\ell,k}}} \right)
\mathbf{U}_{\rm t} \nonumber\\
&=\sum_{\ell=1}^{L}\sum_{s_{\ell}=1}^{S_{\ell}}\beta_{s_{\ell,k}}
\bar{\mathbf{a}}_{\rm r}\left( \phi_{\rm r}^{s_{\ell},k}\right)
\bar{\mathbf{a}}_{\rm t}^{\rm H}\left( {\phi _{\rm{t}}^{{s_\ell,k}}} \right),
\end{align}
where the antenna array response $\bar{\mathbf{a}}_{\rm{t}}\left( \phi_{\rm t}^{s_{\ell},k}\right)\in\mathbb{C}^{N_{\rm t} \times 1}$ of the beamspace channel at the transmitter is given by
\begin{align}
\bar{\mathbf{a}}_{\rm{t}}\left( \phi_{\rm t}^{s_{\ell},k}\right)&=\mathbf{U}_{\rm{t}}^{\rm H}
\mathbf{a}_{\rm{t}}\left( \phi_{\rm t}^{s_{\ell},k}\right) \\\nonumber
&=\left[{\Xi_{{N_{\rm{t}}}}}\left(\phi_{\rm t}^{s_{\ell},k}-\bar\phi_{\rm{t}}^1 \right),\cdots,{\Xi_{{N_{\rm{t}}}}}\left(\phi_{\rm t}^{s_{\ell},k}-\bar\phi_{\rm{t}}^{N_{\rm t}} \right) \right]^{\rm T},
\end{align}
where ${\Xi _N}\left( x \right) = \sum_{n=0}^{N\!-\!1}\frac{1}{N}e^{j2\pi nx}=\frac{{\sin N\pi x}}{{N\sin \pi x}}e^{j\pi x(N\!-\!1)}$.
Similarly, the antenna array response $\bar{\mathbf{a}}_{\rm r}\left( \phi_{\rm r}^{s_{\ell},k}\right)\in\mathbb{C}^{N_{\rm r} \times 1}
$ of the beamspace channel at the receiver is given by
\begin{align}
\bar{\mathbf{a}}_{\rm r}\left( \phi_{\rm r}^{s_{\ell},k}\right)&=\mathbf{U}_{\rm r}^{\rm H}\mathbf{a}_{\rm r}\left( \phi_{\rm r}^{s_{\ell},k}\right)\\\nonumber
&=\left[{\Xi_{{N_{\rm{r}}}}}\left(\phi_{\rm r}^{s_{\ell},k}-\bar\phi_{\rm{r}}^1 \right),\cdots,{\Xi_{{N_{\rm{r}}}}}\left(\phi_{\rm r}^{s_{\ell},k}-\bar\phi_{\rm{r}}^{N_{\rm{r}}} \right) \right]^{\rm T}.
\end{align}

\begin{figure}[t]
	\center{\includegraphics[width=0.5\columnwidth] {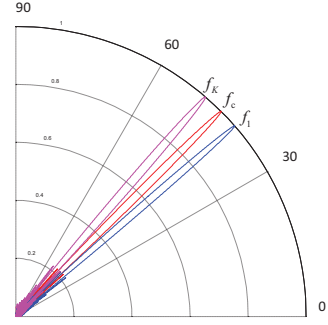}}
	\caption{Power distribution of the beamspace channel at different subcarriers.}
	\label{Fig_beamsquint}
\end{figure}

Observe that ${\Xi _N}\left( x \right)$ has the following characteristics: $\left| {\Xi _N}\left( x \right)\right| \approx 0$ when $|x|\gg 1/N$ \cite{TCOM_WShen_SubspaceCodebook}. Thus $\bar{\mathbf{a}}_{\rm{t}}\left( \phi_{\rm t}^{s_{\ell},k}\right)$ and $\bar{\mathbf{a}}_{\rm r}\left( \phi_{\rm r}^{s_{\ell},k}\right)$ can be regarded as sparse vectors.
As a result, the beamspace channel $\mathbf{H}_{\rm b}[k]$ in (\ref{eq_Hbk2}) is a sparse matrix when the number of clusters $L$ is limited and the AoD/AoA spread of each cluster is small, which are commonly assumed for mmWave channels \cite{Access_TSRappaport_mmwave}.
In other words, the number of focused-energy beams in the beamspace channel is limited. Thus the selection network can select a subset of focused-energy beams, so that the number of RF chains and effective MIMO dimension is reduced without any substantial performance loss.
However, as we have mentioned above, the spatial AoD $\phi_{\rm r}^{s_{\ell},k}$ and spatial AoA $\phi_{\rm t}^{s_{\ell},k}$ are frequency-dependent. Therefore, the beamspace channel $\mathbf{H}_{\rm b}[k]$ is also subcarrier-dependent, where both the values and positions of the non-zero elements of $\mathbf{H}_{\rm b}[k]$  vary with the subcarrier index. This effect of the frequency-dependent beamspace channel is termed as `beam squint' \cite{ICCW_JHBrady_wideband,TWC_XGao_BeamspaceChannelEstimation,TSP_BWang_Spatial}. This is an important feature of the wideband beamspace channel that differentiates it from the traditional narrowband beamspace channel. Fig. \ref{Fig_beamsquint} shows the power distribution of a single-path beamspace channel at different subcarriers with the physical AoD of $\pi/4$. The central carrier frequency is 28 GHz and the system bandwidth is 4 GHz. We observe that the beamspace channel is frequency-dependent.
	
\section{Proposed Beamspace Precoding and Beam Selection}\label{S3}
In this section, we first propose a phase shifter-aided selection network for coping with the effect of beam squint in wideband mmWave MIMO systems. Based on this architecture, a SIC-based beamspace precoding technique is presented for a given beam selection design. Then, a low-complexity energy-max beam selection method is proposed.
	
\subsection{Phase shifter-aided selection network}\label{S3.1}
\begin{figure}[t]
	\center{\includegraphics[width=1\columnwidth] {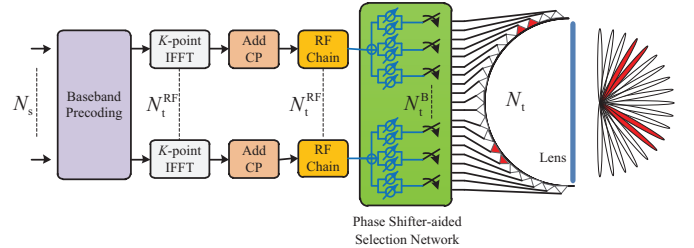}}
	\caption{Transmitter of the proposed phase shifter-aided wideband mmWave MIMO system relying on a lens antenna array. Different from Fig. 1, the traditional selection networks is replaced by the phase shifter-aided selection network.}
	\label{Fig_PS_OFDM_Beamspace_MIMO}
\end{figure}

Note that the beamspace MIMO channel is sparse associated with frequency-dependent non-zero elements (both the locations and values vary over frequencies). However, the selection network realized in the time-domain is frequency-independent, which will lead to power leakage at certain frequencies.
%In order to avoid compromising the performance of some of the subcarriers, several focused-energy beams have to be selected for covering the entire bandwidth, which will result in additional hardware cost and power consumption \cite{ICCW_JHBrady_wideband}.
To address this problem, we propose a phase shifter-aided selection network for mmWave MIMO systems relying on a lens antenna array, as shown in Fig. \ref{Fig_PS_OFDM_Beamspace_MIMO}.

Fig. \ref{Fig_PS_OFDM_Beamspace_MIMO} illustrates the transmitter, while the receiver obeys the inverse architecture. The main difference between the proposed phase shifter-aided selection network and the traditional selection network seen in Fig. \ref{Fig_OFDM_Beamspace_MIMO} is that $N_{\rm t}^{\rm RF}$ RF chains are connected to $N_{\rm t}^{\rm B}$ switches through a sub-array connected phase shifter network. Each switch is associated with two phase shifters and each RF chain is associated with  $\lceil N_{\rm t}^{\rm B}/ N_{\rm t}^{\rm RF} \rceil$ switches. The two-phase-shifter structure is adopted for facilitating the amplitude variations of beamspace precoding \cite{TSP_Zhangxy_variable}, which will be explained in detail later. Within this architecture, the beamspace precoding is realized by the baseband precoding and the sub-array connected phase shifter network.
Through careful design of beam selection and beamspace precoding, the channel's output energy across the entire bandwidth can be captured without increasing the number of RF chains. The design of beam selection and beamspace precoding methods will be discussed  in the following Subsection III-B and Subsection III-C.

The model of the proposed phase shifter-aided mmWave MIMO system can be formulated as a modification of (\ref{eq_yk}). Specifically, the received discrete-time complex baseband signal $\mathbf{y}[k]$ at the $k$-th subcarrier is given by
\begin{align}\label{eq_yk2}
\mathbf{y}[k]=&\sqrt{\rho/N_{\rm s}}\mathbf{W}^{\rm H}_{\rm BB}[k]\mathbf{W}^{\rm H}_{\rm PS}\mathbf{S}^{\rm H}_{\rm r}\mathbf{H}_{\rm b}[k]\mathbf{S}_{\rm t}\mathbf{F}_{\rm PS}\mathbf{F}_{\rm BB}[k]\mathbf{s}[k] \nonumber\\
&+\mathbf{W}^{\rm H}_{\rm BB}[k]\mathbf{W}^{\rm H}_{\rm PS}\mathbf{S}^{\rm H}_{\rm r}\mathbf{n}[k],
\end{align}
where $\mathbf{F}_{\rm PS}\in\mathbb{C}^{N_{\rm t}^{\rm B}\times N_{\rm t}^{\rm RF}}$ is the precoding matrix realized by phase shifters, while $N_{\rm t}^{\rm B}$ is the number of beams selected at the transmitter. Similarly, $\mathbf{W}_{\rm PS}\in\mathbb{C}^{N_{\rm r}^{\rm B}\times N_{\rm r}^{\rm RF}}$ is the combining matrix realized by phase shifters at the receiver, while $N_{\rm r}^{\rm B}$ is the number of beams selected at the receiver. Note that $\mathbf{F}_{\rm PS}$ and $\mathbf{W}_{\rm PS}$ are realized relying on the sub-array connected phase shifters, i.e., each RF chain is associated with $2\lceil N_{\rm t}^{\rm B}/N_{\rm t}^{\rm RF}\rceil$ phase shifters as shown in Fig. \ref{Fig_PS_OFDM_Beamspace_MIMO}. Thus, $\mathbf{F}_{\rm PS}$ and $\mathbf{W}_{\rm PS}$ are block-diagonal matrices. Still referring to (\ref{eq_yk2}),
$\mathbf{S}_{\rm t} \in\mathbb{C}^{N_{\rm t} \times N_{\rm t}^{\rm B}}$ and $\mathbf{S}_{\rm r} \in\mathbb{C}^{N_{\rm r} \times N_{\rm r}^{\rm B}}$ are selection matrices at the transmitter and the receiver, respectively.
The baseband precoding matrix $\mathbf{F}_{\rm BB}[k]$ and the phase shifter-aided precoding matrix $\mathbf{F}_{\rm PS}$ satisfy the transmit power constraint of ${\rm tr}\left(\mathbf{F}^{\rm H}_{\rm BB}[k]\mathbf{F}_{\rm PS}^{\rm H}\mathbf{F}_{\rm PS}\mathbf{F}_{\rm BB}[k] \right)=N_{\rm s}$. The baseband combining matrix $\mathbf{W}_{\rm BB}[k]$ and the phase shifter-aided combining matrix $\mathbf{W}_{\rm PS}$ satisfy the similar power constraint of ${\rm tr}\left(\mathbf{W}^{\rm H}_{\rm BB}[k]\mathbf{W}_{\rm PS}^{\rm H}\mathbf{W}_{\rm PS}\mathbf{W}_{\rm BB}[k] \right)=N_{\rm s}$.
In the following subsections, we will focus our attention on the TPC design, which is composed of beamspace precoding and beam selection.
%The design concept of the TPC can also be used for the receiver combiner design \cite{TWC_AAhmed_LimitedHybridPrecoding}.

\subsection{SIC-based beamspace precoding}\label{S3.2}
To obtain an efficient TPC maximizing the sum-rate, we first decouple the design problems of beam selection and beamspace precoding, based on the fact that the selection matrix is taken from a limited number of candidates.
We consider the throughput optimization problem that captures the fact that the beam selection matrices ${\bf{S}}_{\rm{r}}$ and ${\bf{S}}_{\rm{t}}$ are taken from the sets of candidates $\mathcal{S}_{\rm{r}}$ and $\mathcal{S}_{\rm{t}}$,
\begin{align}\label{eq_Rstar}
\begin{array}{*{20}{l}}
{{R^ \star } = {\mathop {\max }\limits_{{\bf{S}}_{\rm{r}},{\bf{S}}_{\rm{t}},{{\bf{W}}_{{\rm{PS}}}},{{\bf{W}}_{{\rm{BB}}}}[k],{{\bf{F}}_{{\rm{PS}}}},{{\bf{F}}_{{\rm{BB}}}}[k]} R}}\\
{\quad \quad \;\;{\rm{s}}.{\rm{t}}.\quad \begin{array}{*{20}{c}}
	{{{\bf{S}}_{\rm{r}}} \in {\mathcal{S}_{\rm{r}}}}\\
	{{{\bf{S}}_{\rm{t}}} \in {\mathcal{S}_{\rm{t}}}}
	\end{array}.}
\end{array}
\end{align}
The throughput $R$ of the wideband channels is given by \cite{JSAC_FSohrabi_hybrid17}
\begin{align}\label{eq_R}
\begin{array}{l}
R = \frac{1}{K}\sum\limits_{k = 1}^K  {\log _2}\left| {{{\bf{I}}_{N_{\rm{r}}^{\rm{B}}}} + } \right.\frac{\rho }{{{\sigma ^2}{N_{\rm{s}}}}}{\bf{C}}[k]{{{\bf{\tilde H}}}_{\rm{b}}}[k]{{\bf{F}}_{{\rm{PS}}}}{{\bf{F}}_{{\rm{BB}}}}[k]\\
\quad \quad \quad \quad \quad \quad\quad\quad\quad\quad \left. {{\bf{F}}_{{\rm{BB}}}^{\rm{H}}[k]{\bf{F}}_{{\rm{PS}}}^{\rm{H}}{\bf{\tilde H}}_{\rm{b}}^{\rm{H}}[k]} \right|,
\end{array}
\end{align}
where ${\bf{C}}[k] = {{\bf{W}}_{{\rm{PS}}}}{{\bf{W}}_{{\rm{BB}}}}[k]{\left( {{\bf{W}}_{{\rm{BB}}}^{\rm{H}}[k]{\bf{W}}_{{\rm{PS}}}^{\rm{H}}{{\bf{W}}_{{\rm{PS}}}}{{\bf{W}}_{{\rm{BB}}}}[k]} \right)^{ - 1}}\\{\bf{W}}_{{\rm{BB}}}^{\rm{H}}[k]{\bf{W}}_{{\rm{PS}}}^{\rm{H}}$.
The reduced-dimensional beamspace channel $\tilde{\bf{H}}_{\rm{b}}[k]\in\mathbb{C}^{N_{\rm r}^{\rm B}\times N_{\rm t}^{\rm B}}$ after beam selection is defined as
\begin{align}\label{eq_Hbtilde}
\tilde{\bf{H}}_{\rm{b}}[k] = {\bf{S}}_{\rm{r}}^{\rm H}{\bf{ H}_{\rm b}}[k]{{\bf{S}}_{\rm{t}}}.
\end{align}
Since the selection matrices are taken from a limited number of candidates, the throughput optimization problem in (\ref{eq_Rstar}) can be equivalently expressed in the following form
\begin{align}\label{eq_Rstar2}
{R^ \star } = \;\mathop {\max }\limits_{{{\bf{S}}_{\rm{r}}} \in {\mathcal{S}_{\rm{r}}},{{\bf{S}}_{\rm{t}}} \in {\mathcal{S}_{\rm{t}}}} \left\{ {\mathop {\max }\limits_{{{\bf{W}}_{{\rm{PS}}}},{{\bf{W}}_{{\rm{BB}}}}[k],{{\bf{F}}_{{\rm{PS}}}},{{\bf{F}}_{{\rm{BB}}}}[k]} R} \right\}.
\end{align}
The outer maximization is over the legitimate selection matrix candidates (${\mathcal{S}_{\rm{r}}}$ and ${\mathcal{S}_{\rm{t}}}$), while the inner maximization is obtained by finding the beamspace precoding ($\mathbf{F}_{\rm PS}$, $\mathbf{F}_{\rm BB}[k]$) and beamspace combining ($\mathbf{W}_{\rm PS}$, $\mathbf{W}_{\rm BB}[k]$) methods, given the selection matrices ${{\bf{S}}_{\rm{r}}}$ and ${{\bf{S}}_{\rm{t}}}$. In this way, we decouple the design problems of beam selection and beamspace precoding/combining.

However, unfortunately the joint optimization of the throughput over the TPC and receiver combiner is generally intractable \cite{el2013spatially,TCOM_AAlkhateeb_Frequency}. Hence, we follow the common assumption that the receiver combiner is optimal and focus our attention on the TPC design.
The design ideas of TPC given in this paper can be directly used for the receiver combiner design \cite{TWC_AAhmed_LimitedHybridPrecoding}.
Without considering the combining, the throughput tends towards the MI, and we design both ${{\bf{F}}_{{\rm{PS}}}}$ as well as $ {{{\bf{F}}_{{\rm{BB}}}}[k]}$ to maximize the MI ${\cal I}\left( {{{\bf{F}}_{{\rm{PS}}}}, {{{\bf{F}}_{{\rm{BB}}}}[k]} },{\bf{\tilde H}}_{\rm{b}}^{\rm{H}}[k] \right)$, which can be presented as \cite{el2013spatially,TWC_AAhmed_LimitedHybridPrecoding}
\begin{align}\label{eq_I}
&{\cal I}\left( {{{\bf{F}}_{{\rm{PS}}}}, {{{\bf{F}}_{{\rm{BB}}}}[k]} },{\bf{\tilde H}}_{\rm{b}}^{\rm{H}}[k] \right) =\\\nonumber &\frac{1}{K}\!\sum\limits_{k = 1}^K \! {\log _2}\left| {{{\bf{I}}_{N_{\rm{r}}^{\rm{B}}}}\! +\! \frac{\rho }{{{\sigma ^2}{N_{\rm{s}}}}}{{{\bf{\tilde H}}}_{\rm{b}}}[k]{{\bf{F}}_{{\rm{PS}}}}{{\bf{F}}_{{\rm{BB}}}}[k]{\bf{F}}_{{\rm{BB}}}^{\rm{H}}[k]{\bf{F}}_{{\rm{PS}}}^{\rm{H}}{\bf{\tilde H}}_{\rm{b}}^{\rm{H}}[k]} \right|.
\end{align}
The overall problem of MI maximization is modeled as the following outer-inner problem form based on (\ref{eq_Rstar2})
\begin{align}\label{eq_Istar}
{\mathcal{I}^ \star } =\!\!\!\mathop {\max }\limits_{{{\bf{S}}_{\rm{r}}} \in {\mathcal{S}_{\rm{r}}},{{\bf{S}}_{\rm{t}}} \in {\mathcal{S}_{\rm{t}}}} \!\! \left\{ {\mathop {\max }\limits_{{{\bf{F}}_{{\rm{PS}}}},{{\bf{F}}_{{\rm{BB}}}}[k]} \!\!\!{\cal I}\left( {{{\bf{F}}_{{\rm{PS}}}}, {{{\bf{F}}_{{\rm{BB}}}}[k]} },{\bf{\tilde H}}_{\rm{b}}^{\rm{H}}[k] \right)} \right\}.
\end{align}

This subsection focuses on the inner problem formulated in (\ref{eq_Istar}). Our goal is to obtain the beamspace precoding scheme including the phase shifter-aided precoding matrix $\mathbf{F}_{\rm PS}$ and the baseband precoding matrix $\mathbf{F}_{\rm BB}[k]$, that maximizes ${\cal I}\left( {{{\bf{F}}_{{\rm{PS}}}}, {{{\bf{F}}_{{\rm{BB}}}}[k]} },{\bf{\tilde H}}_{\rm{b}}^{\rm{H}}[k] \right)$ for a given ${\bf{S}}_{\rm{t}}$ and ${\bf{S}}_{\rm{r}}$. This problem is solved in two steps. The first step is to design the baseband precoding matrix ${{\bf{F}}_{{\rm{BB}}}}\left[ k \right]$ for a fixed ${{\bf{F}}_{{\rm{PS}}}}$. The optimal baseband precoding matrix is given as  ${{\bf{F}}_{{\rm{BB}}}}\left[ k \right] = {\left( {{\bf{F}}_{{\rm{PS}}}^{\rm{H}}{{\bf{F}}_{{\rm{PS}}}}} \right)^{ - 1/2}}{{\bf{V}}_{{\rm{eff}}}}\left[ k \right] {{\bf{\Gamma }}_{{\rm{eff}}}}\left[ k \right]$ \cite{JSAC_FSohrabi_hybrid17}. Here ${{\bf{V}}_{{\rm{eff}}}}\left[ k \right]\in\mathbb{C}^{N_{\rm{t}}^{\rm RF}\times N_{\rm{s}} }$ gathers the right singular vectors corresponding to the $N_{\rm s}$ largest singular values of ${{\bf{\tilde H}}_{\rm{b}}}\left[ k \right]{{\bf{F}}_{{\rm{PS}}}}{\left( {{\bf{F}}_{{\rm{PS}}}^{\rm{H}}{{\bf{F}}_{{\rm{PS}}}}} \right)^{ - 1/2}}$, and ${{\bf{\Gamma }}_{{\rm{eff}}}}\left[ k \right]\in\mathbb{C}^{N_{\rm{s}}\times N_{\rm{s}} }$ is a diagonal matrix of the power allocated to data streams according to the water filling solution. In the high signal-to-noise ratio (SNR) regime, we have ${{\bf{\Gamma }}_{{\rm{eff}}}}\left[ k \right] \approx {{\bf{I}}_{{N_{\rm{s}}}}}$ \cite{JSAC_FSohrabi_hybrid17}. Note that ${\bf{F}}_{\rm PS}$ is a block-diagonal matrix due to the sub-array connected structure of the proposed phase shifter-aided selection network. Therefore, by normalizing the precoding vector of each sub-array of phase shifters\footnote{As we will discuss later in this subsection, the precoding vector of each sub-array of phase shifters, i.e., the non-zero part of the column vectors of ${\bf{F}}_{\rm PS}$ is a singular vector.}, we have ${\bf{F}}_{{\rm{PS}}}^{\rm{H}}{{\bf{F}}_{{\rm{PS}}}} = \delta{\bf{I}}_{N_{\rm t}^{\rm RF}}$, where $\delta$ is a normalization coefficient ensuring the transmit power constraint of ${\rm tr}\left(\mathbf{F}^{\rm H}_{\rm BB}[k]\mathbf{F}_{\rm PS}^{\rm H}\mathbf{F}_{\rm PS}\mathbf{F}_{\rm BB}[k] \right)=N_{\rm s}$. Substituting ${{\bf{F}}_{{\rm{BB}}}}\left[ k \right] = {\left( {{\bf{F}}_{{\rm{PS}}}^{\rm{H}}{{\bf{F}}_{{\rm{PS}}}}} \right)^{ - 1/2}}{{\bf{V}}_{{\rm{eff}}}}\left[ k \right] {{\bf{\Gamma }}_{{\rm{eff}}}}\left[ k \right] $ into this constraint, we have $\delta=1$. This property of ${\bf{F}}_{\rm PS}$ ensures that the optimal baseband precoding matrix for $N_{\rm{t}}^{\rm RF}= N_{\rm{s}}$ typically satisfies ${\bf F}_{\rm BB}[k]{\bf F}_{\rm BB}^{\rm H}[k]=  {\bf I}_{N_{\rm t}^{\rm RF}}$ in the high-SNR regime\footnote{For wideband mmWave MIMO systems with beam squint, the beams selected for a given subcarrier $k$ may have zero-channel power values. However, the baseband precoding is performed based on the effective channel ${{\bf{\tilde H}}_{\rm{b}}}\left[ k \right]{{\bf{F}}_{{\rm{PS}}}}$, which is of column full rank, since each RF chain at the transmitter is associated with multiple beams. Thus, for wideband systems with beam squint, we still have $\mathbf{F}_{\rm BB}[k] \mathbf{F}_{\rm BB}^{\rm H}[k] = \mathbf{I}_{N_{\rm t}^{\rm RF}}$ in the high-SNR regime.}.

By substituting ${\bf F}_{\rm BB}[k]{\bf F}_{\rm BB}^{\rm H}[k]=  {\bf I}_{N_{\rm t}^{\rm RF}}$ into (\ref{eq_I}), the problem of MI maximization only depends on ${\bf{F}}_{\rm PS}$. Thus, the second step of solving the maximization problem of (\ref{eq_I}) is to design ${\bf{F}}_{\rm PS}$ for maximizing the MI expressed as
\begin{align}\label{eq_IFPS}
&{\cal I}\left( {{{\bf{F}}_{{\rm{PS}}}}},{\bf{\tilde H}}_{\rm{b}}^{\rm{H}}[k] \right)=\\\nonumber& \frac{1}{K}\sum\limits_{k = 1}^K  {\log _2}\left| {{{\bf{I}}_{N_{\rm{s}}}} + \frac{{\rho  }}{{{\sigma ^2}{N_{\rm{s}}}}}{\bf{F}}_{{\rm{PS}}}^{\rm{H}}{\bf{\tilde H}}_{\rm{b}}^{\rm{H}}[k]{{{\bf{\tilde H}}}_{\rm{b}}}[k]{{\bf{F}}_{{\rm{PS}}}}} \right|.
\end{align}
Employing Jensen's inequality, (\ref{eq_IFPS}) can be upper-bounded as \cite{TCOM_WShen_SubspaceCodebook}
\begin{align}\label{eq_IFPS2}
&{\cal I}\left( {{{\bf{F}}_{{\rm{PS}}}}},{\bf{\tilde H}}_{\rm{b}}^{\rm{H}}[k] \right) \le \\\nonumber
&{\log _2}\left| {{{\bf{I}}_{N_{\rm{s}}}} + \frac{{\rho  }}{{{\sigma ^2}{N_{\rm{s}}}}}{\bf{F}}_{{\rm{PS}}}^{\rm{H}}\left( {\frac{1}{K}\sum\limits_{k = 1}^K {{\bf{\tilde H}}_{\rm{b}}^{\rm{H}}[k]{{{\bf{\tilde H}}}_{\rm{b}}}[k]} } \right){{\bf{F}}_{{\rm{PS}}}}} \right|.
\end{align}
Upon defining ${\bf{R}} = \frac{1}{K}\sum\limits_{k = 1}^K {{\bf{\tilde H}}_{\rm{b}}^{\rm{H}}[k]{{{\bf{\tilde H}}}_{\rm{b}}}[k]} $, the desired phase shifter-aided precoding matrix ${\bf{F}}_{{\rm{PS}}}^\star$ maximizing the upper bound of ${\cal I}\left( {{{\bf{F}}_{{\rm{PS}}}}},{\bf{\tilde H}}_{\rm{b}}^{\rm{H}}[k] \right)$ in (\ref{eq_IFPS2}) is given by
\begin{align} \label{eq_I2}
\begin{array}{*{20}{l}}
{{\bf{F}}_{{\rm{PS}}}^ \star  = \mathop {{\rm{arg}}\;{\rm{max}}}\limits_{{{\bf{F}}_{{\rm{PS}}}}} \;\;\;{{\log }_2}\left| {{{\bf{I}}_{{N_{\rm{s}}}}} + \frac{\rho }{{{\sigma ^2}{N_{\rm{s}}}}}{\bf{F}}_{{\rm{PS}}}^{\rm{H}}{\bf{R}}{{\bf{F}}_{{\rm{PS}}}}} \right|}\\
{\;\;\;\quad \quad {\rm{s}}.{\rm{t}}.\;\;\;\left\{ {\begin{array}{*{20}{c}}
		{{\bf{F}}_{{\rm{PS}}}^{\rm{H}}{{\bf{F}}_{{\rm{PS}}}} = {{\bf{I}}_{N_{\rm{t}}^{{\rm{RF}}}}}}\\
		{{{\bf{F}}_{{\rm{PS}}}}\;{\rm{is}}\;{\rm{block-diagonal}}}
		\end{array}.} \right.}
\end{array}
\end{align}
Once ${\bf{F}}_{{\rm{PS}}}^ \star$ has been obtained, the baseband precoding matrix is correspondingly given by ${{\bf{F}}^\star_{{\rm{BB}}}}\left[ k \right] = {\left( {{\bf{F}}^\star_{\rm{PS}}}^{\rm{H}}{{\bf{F}}^\star_{{\rm{PS}}}} \right)^{ - 1/2}}{{\bf{V}}_{{\rm{eff}}}}\left[ k \right] {{\bf{\Gamma }}_{{\rm{eff}}}}\left[ k \right]$.
Next, we present a SIC-based algorithm to solve (\ref{eq_I2}).

Since ${\bf{R}}$ is a Hermitian positive definite matrix, it can be decomposed as ${\bf{R}} = {{\bf{Q}}^{\rm{H}}}{\bf{Q}}$. Then, the optimization target of (\ref{eq_I2}) can be presented as
\begin{align} \label{eq_I3}
&{\log _2}\left| {{{\bf{I}}_{{N_{\rm{s}}}}} + \frac{\rho }{{{\sigma ^2}{N_{\rm{s}}}}}{\bf{F}}_{{\rm{PS}}}^{\rm H}{\bf{R}}{{\bf{F}}_{{\rm{PS}}}}} \right| = \\\nonumber
&{\log _2}\left| {{{\bf{I}}_{{N_{\rm{s}}}}} + \frac{\rho }{{{\sigma ^2}{N_{\rm{s}}}}}{\bf{Q}}{{\bf{F}}_{{\rm{PS}}}}{\bf{F}}_{{\rm{PS}}}^{\rm H}{{\bf{Q}}^{\rm H}}} \right|.
\end{align}
By expressing ${{\bf{F}}_{{\rm{PS}}}}{\rm{ = }}\left[ {{{\bf{F}}_{{\rm{PS,}}{N_{\rm{s}}} -1}},{{\bf{f}}_{{\rm{PS,}}{N_{\rm{s}}}}}} \right]$, where ${{\bf{f}}_{{\rm{PS,}}{N_{\rm{s}}}}}$ is the $N_{\rm s}$-th column vector of ${{\bf{F}}_{{\rm{PS}}}}$ and ${{\bf{F}}_{{\rm{PS,}}{N_{\rm{s}}} - 1}}$ is the submatrix of ${{\bf{F}}_{{\rm{PS}}}}$ by removing ${{\bf{f}}_{{\rm{PS,}}{N_{\rm{s}}}}}$, (\ref{eq_I3}) is rewritten as \cite{JSAC_XGao_EnergyEfficient}
\begin{align}\label{eq_I4}
\begin{array}{*{20}{l}}
{{{\log }_2}\left| {{{\bf{I}}_{{N_{\rm{s}}}}} + \frac{\rho }{{{\sigma ^2}{N_{\rm{s}}}}}{\bf{Q}}{{\bf{F}}_{{\rm{PS}},{N_{\rm{s}}} - 1}}{\bf{F}}_{{\rm{PS}},{N_{\rm{s}}} - 1}^{\rm{H}}{{\bf{Q}}^{\rm{H}}}} \right.}\\
\begin{array}{l}
\quad \quad \left. { + \frac{\rho }{{{\sigma ^2}{N_{\rm{s}}}}}{\bf{Q}}{{\bf{f}}_{{\rm{PS}},{N_{\rm{s}}}}}{\bf{f}}_{{\rm{PS}},{N_{\rm{s}}}}^{\rm{H}}{{\bf{Q}}^{\rm{H}}}} \right|= \\
{\log _2}\left| {{{\bf{T}}_{{N_{\rm{s}}}}}} \right| + {\log _2}\left| {1 + \frac{\rho }{{{\sigma ^2}{N_{\rm{s}}}}}{\bf{f}}_{{\rm{PS}},{N_{\rm{s}}}}^{\rm{H}}{{\bf{Q}}^{\rm{H}}}{\bf{T}}_{{N_{\rm{s}}}}^{ - 1}{\bf{Q}}{{\bf{f}}_{{\rm{PS}},{N_{\rm{s}}}}}} \right|,
\end{array}
\end{array}
\end{align}
where ${{\bf{T}}_{{N_{\rm{s}}}}} = {{\bf{I}}_{{N_{\rm{s}}}}} + \frac{\rho }{{{\sigma ^2}{N_{\rm{s}}}}}{\bf{Q}}{{\bf{F}}_{{\rm{PS}},{N_{\rm{s}}} - 1}}{\bf{F}}_{{\rm{PS}},{N_{\rm{s}}} - 1}^{\rm{H}}{\bf{Q}}^{\rm{H}}$.
Note that $ {\log _2}\left| {{{\bf{T}}_{{N_{\rm{s}}}}}} \right|$ has the same form as the optimization target in (\ref{eq_I3}). Thus, by defining ${{\bf{F}}_{{\rm{PS}},n}}=[{{\bf{F}}_{{\rm{PS}},n - 1}},{{\bf{f}}_{{\rm{PS}},n}}]$, where ${{\bf{f}}_{{\rm{PS,}}{n}}}$ is the $n$-th column vector of ${{\bf{F}}_{{\rm{PS}},n}}$ and  ${{\bf{F}}_{{\rm{PS,}}n - 1}}$ is the submatrix of ${{\bf{F}}_{{\rm{PS}},n}}$ by removing ${{\bf{f}}_{{\rm{PS}},n}}$ ($n=1,2,\cdots,N_{\rm s}$), the optimization target in (\ref{eq_I3}) can be further decomposed as
\begin{align}\label{eq_I5}
&{\log _2}\left| {{{\bf{I}}_{{N_{\rm{s}}}}} + \frac{\rho }{{{\sigma ^2}{N_{\rm{s}}}}}{\bf{F}}_{{\rm{PS}}}^{\rm H}{\bf{R}}{{\bf{F}}_{{\rm{PS}}}}} \right| =\\\nonumber &\sum\limits_{n = 1}^{{N_{\rm{s}}}} {{{\log }_2}\left( {1 + \frac{\rho }{{{\sigma ^2}{N_{\rm{s}}}}}{\bf{f}}_{{\rm{PS,}}n}^{\rm H}{{\bf{Q}}^{\rm H}}{\bf{{\rm\bf T}}}_n^{ - 1}{\bf{Q}}{{\bf{f}}_{{\rm{PS,}}n}}} \right)} ,
\end{align}
where ${{\bf{T}}_n} = {{\bf{I}}_{{N_{\rm{s}}}}} + \frac{\rho }{{{\sigma ^2}{N_{\rm{s}}}}}{\bf{Q}}{{\bf{F}}_{{\rm{PS}},n - 1}}{\bf{F}}_{{\rm{PS}},n - 1}^{\rm{H}}{\bf{Q}}^{\rm{H}}$ and ${{\bf{{\rm \bf T}}}_1} = {{\bf{I}}_{{N_{\rm{s}}}}}$.
Thus, the optimization problem of (\ref{eq_I2}) can be decomposed into ${N_{\rm{s}}}$ subproblems, where the $n$-th subproblem is
\begin{align}\label{eq_In}
\begin{array}{*{20}{l}}
{{\bf{f}}_{{\rm{PS}},n}^ \star {\rm{ = }}\mathop {\arg \;\max }\limits_{{{\bf{f}}_{{\rm{PS}},n}}} \;\;\;{\kern 1pt} {{\log }_2}\left( {1 + \frac{\rho }{{{\sigma ^2}{N_{\rm{s}}}}}{\bf{f}}_{{\rm{PS}},n}^{\rm{H}}{{\bf{G}}_n}{{\bf{f}}_{{\rm{PS}},n}}} \right)}\\
{\;\;\;{\kern 1pt} \;\;\;{\kern 1pt} \;\quad {\rm{s}}.{\rm{t}}.\;\;\;{\kern 1pt} \;\;\;{\kern 1pt} {\bf{f}}_{{\rm{PS}},n}^{\rm{H}}{{\bf{f}}_{{\rm{PS}},n}} = 1,}
\end{array}
\end{align}
where ${{\bf{G}}_n} = {{\bf{Q}}^{\rm H}}{\bf{{\rm\bf T}}}_n^{ - 1}{\bf{Q}}$.

Inspired by the SIC concept \cite{JSAC_XGao_EnergyEfficient}, we propose to successively solve the above $N_{\rm s}$ subproblems. The proposed SIC-based beamspace precoding scheme is summarized in Algorithm \ref{alg_1} and explained as follows. The algorithm starts by solving the first subproblem (\ref{eq_In}). The first column of ${\bf{F}}_{{\rm{PS}}}^ \star$ is obtained as ${\bf{f}}_{{\rm{PS}},1}^ \star$. Then, it is used to update the matrices $\mathbf{T}_2={{\bf{I}}_{{N_{\rm{s}}}}} + \frac{\rho }{{{\sigma ^2}{N_{\rm{s}}}}}{\bf{Q}}{{\bf{F}}_{{\rm{PS}},2 - 1}}{\bf{F}}_{{\rm{PS}},2 - 1}^{\rm{H}}{\bf{Q}}^{\rm{H}}$ and ${{\bf{G}}_2} = {{\bf{Q}}^{\rm H}}{\bf{{\rm\bf T}}}_2^{ - 1}{\bf{Q}}$, where ${{\bf{F}}_{{\rm{PS}},2 - 1}}={\bf{f}}_{{\rm{PS}},1}^ \star$. The second column ${\bf{f}}_{{\rm{PS}},2}^ \star$ of ${\bf{F}}_{{\rm{PS}}}^ \star$ is obtained by solving the second subproblem. We then repeat this procedure until the desired precoding matrix ${{\bf{F}}^ \star_{{\rm{PS}}}}$ is obtained. Then, the baseband precoding matrix is designed as ${{\bf{F}}^\star_{{\rm{BB}}}}\left[ k \right] = {\left( {{\bf{F}}^\star_{\rm{PS}}}^{\rm{H}}{{\bf{F}}^\star_{{\rm{PS}}}} \right)^{ - 1/2}}{{\bf{V}}_{{\rm{eff}}}}\left[ k \right] {{\bf{\Gamma }}_{{\rm{eff}}}}\left[ k \right]$.
\begin{algorithm}[tb!]
	\renewcommand{\algorithmicrequire}{\textbf{Input:}}
	\renewcommand\algorithmicensure {\textbf{Output:} }
	\caption{Proposed SIC-based beamspace precoding}
	\label{alg_1}
	\begin{algorithmic}[1]
		\STATE\textbf{Input:} \\
		$\mathbf{T}_1$, $\mathbf{G}_1$,
		${\bf{F}}_{{\rm{PS}},{0}}=\emptyset$\\
		\FOR {$1\le n\le N_{\rm s}$}
		\STATE Compute ${\bf{f}}_{{\rm{PS}},n}^ \star$ by solving (\ref{eq_In})
		\STATE ${{\bf{F}}_{{\rm{PS}},n}}{\rm{ = }}\left[ {{{\bf{F}}_{{\rm{PS,}}{n -1}},{{\bf{f}}^ \star_{{\rm{PS,}}{n}}}}} \right]$
		\STATE Update $\mathbf{T}_{n+1}$ and $\mathbf{G}_{n+1}$
		\ENDFOR
		\STATE\textbf{Output:}\\
		\STATE ${\bf F}_{{\rm{PS}}}^ \star={\bf F}_{{\rm{PS}},N_{\rm s}}$
		\STATE ${{\bf{F}}^\star_{{\rm{BB}}}}\left[ k \right] = {\left( {{\bf{F}}^\star_{\rm{PS}}}^{\rm{H}}{{\bf{F}}^\star_{{\rm{PS}}}} \right)^{ - 1/2}}{{\bf{V}}_{{\rm{eff}}}}\left[ k \right] {{\bf{\Gamma }}_{{\rm{eff}}}}\left[ k \right]$
	\end{algorithmic}
\end{algorithm}

Finally, we discuss how to solve (\ref{eq_In}) for obtaining ${\bf{f}}_{{\rm{PS}},n}^ \star$.
Defining a vector ${{\bf{\bar f}}_{{\rm{PS,}}n}}$, which gathers the non-zero elements of ${{\bf{f}}_{{\rm{PS,}}n}}$, (\ref{eq_In}) can be rewritten as
\begin{align}\label{eq_In2}
\begin{array}{*{20}{l}}
{{\bf{\bar f}}_{{\rm{PS}},n}^ \star  = \mathop {\arg \;\max }\limits_{{{{\bf{\bar f}}}_{{\rm{PS}},n}}} \quad {{\log }_2}\left( {1 + \frac{\rho }{{{\sigma ^2}{N_{\rm{s}}}}}{\bf{\bar f}}_{{\rm{PS}},n}^{\rm{H}}{{{\bf{\bar G}}}_n}{{{\bf{\bar f}}}_{{\rm{PS}},n}}} \right)}\\
{\quad \quad \quad \;{\rm{s}}.{\rm{t}}.\quad \quad {\bf{\bar f}}_{{\rm{PS}},n}^{\rm{H}}{{{\bf{\bar f}}}_{{\rm{PS}},n}} = 1,}
\end{array}
\end{align}
where ${{\bf{\bar G}}_n}$ is the sub-matrix composed of the elements of ${{\bf{G}}_n}$ corresponding to the non-zero elements of ${{\bf{f}}_{{\rm{PS,}}n}}$.
The solution of the subproblem (\ref{eq_In2}) is given by the first right singular vector $\mathbf{v}_n$ corresponding to the largest singular value of ${{\bf{\bar G}}_n}$ \cite{JSAC_XGao_EnergyEfficient}.
Note the phase shifter-aided precoding matrix is realized by a sub-array connected phase shifter network and each element of $\mathbf{v}_n$ is realized by a pair of phase shifters, as shown in Fig. 3. Given that the amplitude of each element of $\mathbf{v}_n$ is not larger than 1, since $\mathbf{v}_n^{\rm H}\mathbf{v}_n=1$, we can assume that each element of $\mathbf{v}_n$ is expressed as $\alpha{e^{j\beta}}$ ($\alpha \le1$). Our target is to design two phase shifters with the phases of $\beta_1$ and $\beta_2$ to satisfy ${e^{j{\beta _1}}} + {e^{j{\beta _2}}} = \alpha {e^{j\beta }}$, which is equivalently expressed as
\begin{align}\label{eq_cos}
\left\{ \begin{array}{l}
\cos \left( {{\beta _1} - \beta } \right) + \cos \left( {{\beta _2} - \beta } \right) = \alpha \\
\sin \left( {{\beta _1} - \beta } \right) + \sin \left( {{\beta _2} - \beta } \right) = 0
\end{array} \right. .
\end{align}
Solving (\ref{eq_cos}), we have ${\beta _1} = {\cos ^{ - 1}}\left( {\alpha /2} \right) + \beta $ and ${\beta _2} = \beta  - {\cos ^{ - 1}}\left( {\alpha /2} \right)$.

We also estimate the computational complexity of the proposed \textbf{Algorithm 1}. Specifically, to obtain the phase shifter-aided precoding matrix $\mathbf{F}_{\rm PS}$, we have to iteratively compute ${{\bf{G}}_n} = {{\bf{Q}}^{\rm H}}{\bf{{\rm\bf T}}}_n^{ - 1}{\bf{Q}}$ and the SVD of $\bar{\mathbf{G}}_n$ ($n=1,2,\cdots, N_{\rm s}$). The complexity of computing ${{\bf{G}}_n}$ is $\mathcal{O}\left(4{N_{\rm t}^{\rm B}}^3+2n{N_{\rm t}^{\rm B}}^2 \right) $ and that of computing the SVD of ${\bar{{\bf{G}}}_n}$ is $\mathcal{O}\left(\left( {N_{\rm t}^{\rm B}}/{N_{\rm t}^{\rm RF}}\right) ^3 \right) $. Therefore, the complexity of computing $\mathbf{F}_{\rm PS}$ is $\mathcal{O}\left(4N_{\rm s}{N_{\rm t}^{\rm B}}^3+N_{\rm s}(N_{\rm s}+1){N_{\rm t}^{\rm B}}^2+N_{\rm s}\left( {N_{\rm t}^{\rm B}}/{N_{\rm t}^{\rm RF}}\right) ^3 \right) $. The baseband precoding matrix $\mathbf{F}_{\rm BB}[k]$ is computed by step 9, which requires the computational complexity of $\mathcal{O}\left({N_{\rm r}^{\rm B}}{N_{\rm t}^{\rm RF}}^2\right) $. To sum up, the complexity of our proposed SIC-based beamspace precoding is  $\mathcal{O}\left(4N_{\rm s}{N_{\rm t}^{\rm B}}^3+N_{\rm s}(N_{\rm s}+1){N_{\rm t}^{\rm B}}^2 +N_{\rm s}\left( {N_{\rm t}^{\rm B}}/{N_{\rm t}^{\rm RF}}\right) ^3 \right.$ \\
	$\left.+K{N_{\rm r}^{\rm B}}{N_{\rm t}^{\rm RF}}^2 \right) $.

\subsection{Energy-max beam selection}\label{S3.3}
This subsection focuses on the design problem of beam selection. The inner maximization problem in (\ref{eq_Istar}) is solved by the proposed SIC-based beamspace precoding scheme given the beam selection matrices $\mathbf{S}_{\rm t}$ and $\mathbf{S}_{\rm r}$. Then, the overall solution to the MI maximization problem in (\ref{eq_Istar}) can be obtained by solving the outer maximization problem through an exhaustive search over $\mathcal{S}_{\rm t}$ and $\mathcal{S}_{\rm r}$. Since the exhaustive search is of high complexity, in this subsection, we develop a low-complexity beam selection method.

%Substituting (\ref{eq_Hbtilde}) in to (\ref{eq_yk2}),
%\begin{align}\label{eq_yk2}
%\mathbf{y}[k]=&\sqrt{\rho/N_{\rm s}}\mathbf{W}^{\rm H}_{\rm BB}[k]\mathbf{W}^{\rm H}_{\rm PS}\tilde{\bf{H}}_{\rm{b}}[k]\mathbf{F}_{\rm PS}\mathbf{F}_{\rm BB}[k]\mathbf{s}[k] \nonumber\\
%&+\mathbf{W}^{\rm H}_{\rm BB}[k]\mathbf{W}^{\rm H}_{\rm PS}\tilde{\mathbf{n}}[k],
%\end{align}
%where $\tilde{\mathbf{n}}[k]=\mathbf{S}^{\rm H}_{\rm r}\mathbf{n}[k]$.

The beamspace precoding scheme ($\mathbf{F}^\star_{\rm PS}$, $\mathbf{F}^\star_{\rm BB}[k]$) obtained through the inner problem maximization in (\ref{eq_Istar}) only depends on the reduced-dimensional beamspace channel ${\bf{\tilde H}}_{\rm{b}}^{\rm{H}}[k]$. Thus, the MI optimization problem (\ref{eq_Istar}) is rewritten as
\begin{align}\label{eq_Istar2}
{{\cal I}^ \star } = \;\mathop {\max }\limits_{{{\bf{S}}_{\rm{r}}} \in {\mathcal{S}_{\rm{r}}},{{\bf{S}}_{\rm{t}}} \in {\mathcal{S}_{\rm{t}}}} {\cal I}\left( {{\bf{\tilde H}}_{\rm{b}}^{\rm{H}}[k]}\right).
\end{align}
By using (\ref{eq_IFPS}), we have
\begin{align}\label{eq_IHb}
{\cal I}\left( {{\bf{\tilde H}}_{\rm{b}}^{\rm{H}}[k]} \right) {\rm{ = }}\frac{1}{K}\sum\limits_{k = 1}^K  {\log _2}\left| {{{\bf{I}}_{N_{\rm{s}}}} + \frac{{\rho  }}{{{\sigma ^2}{N_{\rm{s}}}}}{{\bf{F}}^\star_{\rm{PS}}}^{\rm{H}}{\bf{\tilde H}}_{\rm{b}}^{\rm{H}}[k]{{{\bf{\tilde H}}}_{\rm{b}}}[k]{{\bf{F}}^\star_{{\rm{PS}}}}} \right|.
\end{align}
Similar to the decoupling of the beamspace precoding and combining design, we focus our attention on the beam selection design at the transmitter and temporarily assume that ${{\bf{S}}_{\rm{r}}}{\bf{S}}_{\rm{r}}^{\rm{H}} = {{\bf{I}}_{{N_{\rm{r}}}}}$. The design method of the transmit beam selection can be directly used for the receive beam selection.
Thus, the MI in (\ref{eq_IHb}) is rewritten as
\begin{align}\label{eq_ISt}
\begin{array}{l}
{\cal I}\left( {{{\bf{S}}_{\rm{t}}}} \right) = \frac{1}{K}\sum\limits_{k = 1}^K {{{\log }_2}} \left| {{{\bf{I}}_{{N_{\rm{s}}}}} + \frac{\rho }{{{\sigma ^2}{N_{\rm{s}}}}}{\bf{F}}{{_{{\rm{PS}}}^ \star }^{\rm{H}}}{\bf{S}}_{\rm{t}}^{\rm{H}}{\bf{H}}_{\rm{b}}^{\rm{H}}[k]{{\bf{H}}_{\rm{b}}}[k]{{\bf{S}}_{\rm{t}}}{\bf{F}}_{{\rm{PS}}}^ \star } \right|\\
\quad \quad \;\;\, \overset{(a)}{=} \frac{1}{K}\sum\limits_{k = 1}^K {\left\{ {\sum\limits_{s = 1}^{{N_{\rm{s}}}} {{{\log }_2}\left( {1 + \frac{\rho }{{{\sigma ^2}{N_{\rm{S}}}}}\lambda _s[k]^2} \right)} } \right\}} ,
\end{array}
\end{align}
where (a) is obtained by the singular value decomposition (SVD) of the matrix ${{\bf{H}}_{\rm{b}}}[k]{{\bf{S}}_{\rm{t}}}{\bf{F}}_{{\rm{PS}}}^ \star$ and $\left\{ {{\lambda _s}}[k] \right\}_{s = 1}^{{N_s}}$ are the singular values.

By employing Jensen's inequality, (\ref{eq_ISt}) is upper-bounded as
\begin{align}\label{eq_ISt2}
\begin{array}{l}
I\left( {{{\bf{S}}_{\rm{t}}}} \right) \le \frac{{{N_{\rm{s}}}}}{K}\sum\limits_{k = 1}^K {\left\{ {{{\log }_2}\left( {1 + \frac{\rho }{{{\sigma ^2}N_{\rm{s}}^2}}\sum\limits_{s = 1}^{{N_{\rm{s}}}} {\lambda _s^2[k]} } \right)} \right\}} \\
\quad \quad \;\;\; \le {N_{\rm{s}}}{\log _2}\left( {1 + \frac{\rho }{{{\sigma ^2}N_{\rm{s}}^2K}}\sum\limits_{k = 1}^K {{\sum\limits_{s = 1}^{{N_{\rm{s}}}} {\lambda _s^2[k]} }} } \right).
\end{array}
\end{align}
The transmit beam selection matrix ${{{\bf{S}}_{\rm{t}}}}$ is designed for maximizing the upper bound of the MI \cite{JSAC_FSohrabi_hybrid17}.
Due to the fact that $\sum\limits_{s = 1}^{{N_{\rm{s}}}} {\lambda _s^2\left[ k \right]}  = \left\| {{{\bf{H}}_{\rm{b}}}[k]{{\bf{S}}_{\rm{t}}}{\bf{F}}_{{\rm{PS}}}^ \star } \right\|_{\rm F}^2$, the selection matrix of the transmitter is designed as
\begin{align}\label{eq_St}
{\bf{S}}_{\rm{t}}^ \star  = \mathop {\arg \max }\limits_{{{\bf{S}}_{\rm{t}}}} \sum\limits_{k = 1}^K\left\| {{{\bf{H}}_{\rm{b}}[k]}{{\bf{S}}_{\rm{t}}}{\bf{F}}_{{\rm{PS}}}^ \star } \right\|_{\rm F}^2.
\end{align}

We observe from (\ref{eq_St}) that to maximize the upper bound of the MI, the selection matrix ${{\bf{S}}_{\rm{t}}}$ should select the specific transmit beams (i.e., columns of the beamspace channel ${{\bf{H}}_{\rm{b}}[k]}$), which are then combined by ${\bf{F}}_{{\rm{PS}}}^ \star $ to capture as much of the channel's output energy as possible.
To solve (\ref{eq_St}), we first define
\begin{align}
{\tilde{\bf{S}}_{\rm{t}}}={{\bf{S}}_{\rm{t}}}{\bf{F}}_{{\rm{PS}}}^ \star .
\end{align}
Since ${\bf{F}}_{{\rm{PS}}}^ \star$ is a block-diagonal matrix, ${\tilde{\bf{S}}_{\rm{t}}}\in\mathbb{C}^{N_{\rm t}\times N_{\rm s}}$ is a matrix composed of $N_{\rm s}$ columns, each of which has $\lceil N_{\rm t}^{\rm B}/N_{\rm s}\rceil$ non-zero elements. This means that each column of ${\tilde{\bf{S}}_{\rm{t}}}$ can select $\lceil N_{\rm t}^{\rm B}/N_{\rm s}\rceil$ transmit beams and combine them with different weights, which are determined by the diagonal block of ${\bf{F}}_{{\rm{PS}}}^ \star$. Then we rewrite the optimization target in (\ref{eq_St}) as
\begin{align}
\sum\limits_{k = 1}^K\left\| {{{\bf{H}}_{\rm{b}}[k]}{{\bf{S}}_{\rm{t}}}{\bf{F}}_{{\rm{PS}}}^ \star } \right\|_{\rm F}^2 = {\rm{tr}}\left({\tilde{\bf{S}}_{\rm{t}}^{\rm{H}}\left( \sum\limits_{k = 1}^K{\bf{H}}_{\rm{b}}^{\rm{H}}[k]{{\bf{H}}_{\rm{b}}}[k]\right) {\tilde{\bf{S}}}_{\rm{t}}} \right) .
\end{align}
Observe that $\sum\limits_{k = 1}^K {{\bf{H}}_{\rm{b}}^{\rm{H}}\left[ k \right]}{{\bf{H}}_{\rm{b}}}\left[ k \right] $ is a diagonal-dominant matrix \cite{TCOM_WShen_SubspaceCodebook}. The $i$-th diagonal element of $\frac{1}{K}\sum\limits_{k = 1}^K {{\bf{H}}_{\rm{b}}^{\rm{H}}\left[ k \right]}{{\bf{H}}_{\rm{b}}}\left[ k \right] $ represents the energy of the $i$-th transmit beam of the beamspace channel averaged over the entire bandwidth. Thus, if ${\tilde{\bf{S}}}_{\rm{t}}$ is going to select $N_{\rm t}^{\rm B}$ transmit beams and combine them as $N_{\rm s}$ beams to transmit signals,  ${\bf{S}}_{\rm{t}}^ \star$ should be designed to select the $N_{\rm t}^{\rm B}$ largest diagonal elements\footnote{The transmit beams associated with the $N_{\rm t}^{\rm B}$ largest diagonal elements of $\frac{1}{K}\sum\limits_{k = 1}^K {{\bf{H}}_{\rm{b}}^{\rm{H}}\left[ k \right] {{\bf{H}}_{\rm{b}}}\left[ k \right]}$ are selected as candidate beams. Then, the indices of the candidate beams are sorted in an ascending order. Thus, the locations of non-zero elements in the columns of ${\bf{S}}_{\rm{t}}^ \star$ are in an ascending order.} of $\frac{1}{K}\sum\limits_{k = 1}^K {{\bf{H}}_{\rm{b}}^{\rm{H}}\left[ k \right] {{\bf{H}}_{\rm{b}}}\left[ k \right]} $.

\begin{figure}[t]
	\center{\includegraphics[width=1\columnwidth] {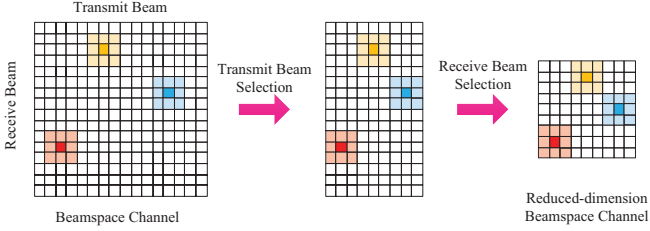}}
	\caption{Illustration of the proposed energy-max beam selection method. The transmitter selects the focused-energy transmit beams and the receiver selects the focused-energy receive beams to form the reduced-dimensional beamspace channel.}
	\vspace{-0mm}
	\label{Fig_Beam_selection}
\end{figure}

Similar to the transmitter,  the selection matrix ${\bf{S}}_{\rm{r}}^ \star$ at the receiver is designed to select the $N_{\rm r}^{\rm B}$ largest diagonal elements of $\frac{1}{K}\sum\limits_{k = 1}^K {{{\bf{H}}_{\rm{b}}}\left[ k \right]{\bf S}^{\star}_{\rm t}{{\bf S}^{\star}_{\rm t}}^{\rm H} {\bf{H}}^{\rm{H}}_{\rm{b}}\left[ k \right]} $ to capture as much of the energy of the receive beams of the beamspace channel as possible. The procedure of the energy-max beam selection method is shown in Fig. \ref{Fig_Beam_selection}. The transmitter first selects $N_{\rm t}^{\rm B}$ focused-energy transmit beams corresponding to the $N_{\rm t}^{\rm B}$ largest diagonal elements of $\frac{1}{K}\sum\limits_{k = 1}^K {{\bf{H}}_{\rm{b}}^{\rm{H}}\left[ k \right]}{{\bf{H}}_{\rm{b}}}\left[ k \right] $. The selection matrix at the transmitter is denoted by ${\bf{S}}_{\rm{t}}^ \star$. Then the receiver selects $N_{\rm r}^{\rm B}$ focused-energy receive beams corresponding to the $N_{\rm r}^{\rm B}$ largest diagonal elements of $\frac{1}{K}\sum\limits_{k = 1}^K {{{\bf{H}}_{\rm{b}}}\left[ k \right]{\bf S}^{\star}_{\rm t}{{\bf S}^{\star}_{\rm t}}^{\rm H} {\bf{H}}^{\rm{H}}_{\rm{b}}\left[ k \right]} $. The selection matrix at the receiver is denoted by ${\bf{S}}_{\rm{r}}^ \star$. Thus, the reduced-dimensional beamspace channel is obtained by $\tilde{\bf{H}}_{\rm{b}}[k] = {{\bf{S}}^\star_{\rm{r}}}^{\rm H}{\bf{ H}_{\rm b}}[k]{{\bf{S}}^ \star_{\rm{t}}}$.

Naturally, having more beams selected captures more energy of the beamspace channel. Ideally, setting $N_{\rm t}^{\rm B}=N_{\rm t}$ and $N_{\rm r}^{\rm B}=N_{\rm r}$ enables all the beamspace channel energy to be captured. However, increasing $N_{\rm t}^{\rm B}$ and $N_{\rm r}^{\rm B}$ leads to higher hardware cost and higher power consumption.
Given the sparsity of the beamspace channel, a small number of transmit beams and receive beams are capable of capturing most of the channel's output energy. It is expected that a better trade-off between the hardware cost/power consumption and the sum-rate performance can be achieved by appropriately choosing the values of $N_{\rm t}^{\rm B}$ and $N_{\rm r}^{\rm B}$. It is proved in Appendix that $N_{\rm t}^{\rm B}$ and $N_{\rm r}^{\rm B}$ should be set as
\begin{align} \label{eq_NtB}
N_{\rm{t}}^{\rm{B}} = \left\lceil {\frac{L{{N_{\rm{t}}}B}}{{2{f_{\rm c}}}}} \right\rceil,
N_{\rm{r}}^{\rm{B}} = \left\lceil {\frac{L{{N_{\rm{r}}}B}}{{2{f_{\rm c}}}}} \right\rceil.
\end{align}
We observe that the number of required beams increases with the system bandwidth. This is because the effect of beam squint leads to heavier power leakage for a high bandwidth.

\section{Simulation Results}\label{S4}
In this section, we evaluate the performance of the proposed phase shifter-aided selection network as well as the associated TPC design through numerical simulations. The key system parameters are summarized in Table \ref{Table_1}. The delays of the dominant paths are assumed to follow the uniform distribution within $[0,20]$(ns)\cite{TSP_GaoX_wideband}. Each dominant path is contributed by 20 subpaths, whose delay offsets are within $[-0.1,0.1]$(ns). The path gain $\alpha_{s_{\ell}}$ follows the complex Gaussian distribution $\mathcal{CN}(0,1)$. The AoAs/AoDs of dominant paths are assumed to follow the uniform distribution within $[-1/2\pi,1/2\pi]$. The maximum AoA/AoD spread of dominant paths is $5^\circ$\cite{3GPPTR_SCM}. We simulate the MI ${\cal I}^\star$ defined in (\ref{eq_Istar}), which is achieved via the proposed SIC-based beamspace precoding and energy-max beam selection. We also present the energy efficiency (EE) defined as $\eta={\cal I}^\star/P_{\rm total}$, where $P_{\rm total}$ is the total energy consumption, defined as $P_{\rm total}=\rho + P_{\rm c}+N_{\rm t}^{\rm RF}P_{\rm RF} + 2N_{\rm t}^{\rm B}P_{\rm PS} + N_{\rm t}^{\rm B}P_{\rm Switch}$. Here, $P_{\rm RF}$, $P_{\rm PS}$ and $P_{\rm Switch}$ are the energy consumed by each RF chain, phase shifter and switch, respectively. We adopt the typical values of $P_{\rm RF}=250 $ mW, $P_{\rm PS}=10 $ mW, and $P_{\rm Switch}=5$ mW \cite{Access_RMendez_hybridmimo}. $P_{\rm c}=p_{\rm c}\times N_{\rm c}$ is the computational cost with $p_{\rm c}=14.1$ mW/MOps \cite{TSP_YLee_ahybrid} being the energy consumption per $10^6$ elementary operations (MOps) of the digital signal processor (DSP) and $N_{\rm c}$ being the computational complexity as discussed in Section III.
	\begin{table}[tb!]
	\begin{center}
		\caption{System parameters for simulation}  \label{Table_1}
		\begin{tabular}{|l|r|}
			\hline
			{\textbf{Parameter}} & \textbf{Values} \\
			\hline
			\makecell[l]{Central carrier frequency $f_{\rm c}$ (GHz) } & 28 \\
			\hline
			\tabincell{c}{System's bandwidth $B$ (GHz) } & {$1 \sim 4$} \\
			\hline
			\tabincell{c}{$\#$ of dominant channel paths $L$} & 10 \\
			\hline
			\tabincell{c}{$\#$ of subpaths of each cluster $S_{\ell}$} & {20} \\
			\hline
			\tabincell{c}{$\#$ of subcarriers $K$} & {1024} \\
			\hline
			\tabincell{c}{$\#$ of TAs and RAs $N_{\rm t}=N_{\rm r}$ } &  $ 64$ \\
			\hline
			\tabincell{c}{$\#$ of data streams $N_{\rm s}$ } &  $ 4 \sim 10$ \\
			\hline
			\tabincell{c}{$\#$ of RF chains $N_{\rm t}^{\rm RF}=N_{\rm r}^{\rm RF}$} &  $ 4 \sim 10$ \\
			\hline
			\tabincell{c}{Transmit power $\rho$ (W)} &  $ 0.2 \sim 2$ \\
			\hline
		\end{tabular}
	\end{center}
\end{table}

For comparison, we also simulate the MI of the traditional wideband mmWave MIMO systems relying on the SVD-based TPC, which is defined as
\begin{align}\label{eq_IT}
{\cal I}_{\rm SVD} \!=\!\! \frac{1}{K}\!\sum\limits_{k = 1}^K  {\log _2}\left| {{{\bf{I}}_{N_{\rm{r}}^{\rm{RF}}}}\! +\!  \frac{\rho }{{{\sigma ^2}{N_{\rm{s}}}}}{{{\bf{\tilde H}}}_{\rm{b}}}[k]{{\bf{F}}_{{\rm{BB}}}}[k]{\bf{F}}_{{\rm{BB}}}^{\rm{H}}[k]{\bf{\tilde H}}_{\rm{b}}^{\rm{H}}[k]} \right|,
\end{align}
where the SVD-based baseband precoding matrix ${\bf{F}}_{{\rm{BB}}}[k]$ is composed of the right singular vectors of ${{{\bf{\tilde H}}}_{\rm{b}}}[k]$ corresponding to the ${N_{\rm{r}}^{\rm{RF}}}$ largest singular values \cite{JSAC_FSohrabi_hybrid17}.
The EE of the traditional SVD-based wideband mmWave MIMO systems is defined as $\eta_{\rm SVD}={{\cal I}_{\rm SVD}}/( \rho +p_{\rm c}N_{\rm SVD}+ N_{\rm t}^{\rm RF}P_{\rm RF}+N_{\rm t}^{\rm RF}P_{\rm Switch}) $, where $N_{\rm SVD}$ is the computational complexity of the SVD-based TPC. For fair comparison, we adopt the low-complexity energy-max beam selection for both systems.

\begin{figure}[tb!]
	\vspace{-0mm}
	\center{\includegraphics[width=0.95\columnwidth]{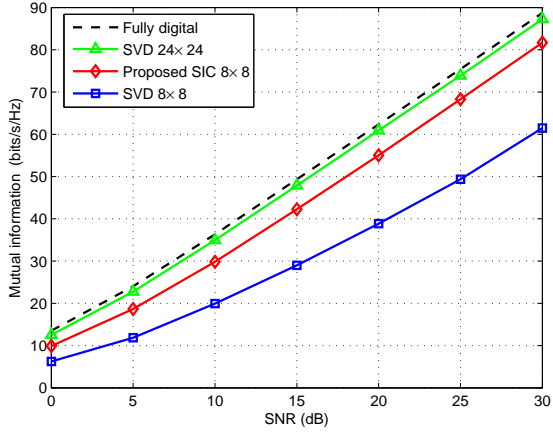}}
	\vspace{-3mm}
	\caption{MI against the SNR. The proposed SIC-based method outperforms the traditional SVD-based method when the same number of RF chains is considered.}
	\label{Fig_SE_SNR}
\end{figure}
In Fig. \ref{Fig_SE_SNR}, we present the MI vs the SNR, which is defined as $\rho/\sigma^2$. The system's bandwidth is $B=2$ GHz. The number of RF chains is $N_{\rm t}^{\rm RF}=N_{\rm r}^{\rm RF}=8$ for the proposed SIC-based method, which is denoted by ``SIC $8 \times 8$''. The number of data streams is set to $N_{\rm s}=N_{\rm t}^{\rm RF}=N_{\rm r}^{\rm RF}$. For the given system parameters, the number of selected beams should be set as $N_{\rm t}^{\rm B}=N_{\rm r}^{\rm B}=24$ according to (\ref{eq_NtB}). The traditional SVD-based methods using 8 RF chains and 24 RF chains are denoted by ``SVD $8 \times 8$'' and ``SVD $24 \times 24$'', respectively. We also present the fully digital precoding method using $64$ RF chains at the transmitter and receiver as the benchmark. We observe that the proposed SIC-based method outperforms the traditional SVD-based methods, when the same number of RF chains is considered. As expected, the ``SIC $8 \times 8$'' method has worse performance than the ``SVD $24 \times 24$'' method, since the SIC-based method combines the selected 24 beams into 8 beams through a normalized precoding matrix $\mathbf{F}_{\rm PS}$. Thus, the array beamforming gain in the most desired direction is lower than that of the traditional ``SVD $24 \times 24$'' method. However, the traditional ``SVD $24 \times 24$'' method requires more RF chains, which will lead to higher hardware cost and higher energy consumption, as it will be verified in the following Fig. \ref{Fig_EE_Pt}. 	

\begin{figure}[tb!]
	\vspace{-0mm}
	\center{\includegraphics[width=0.95\columnwidth]{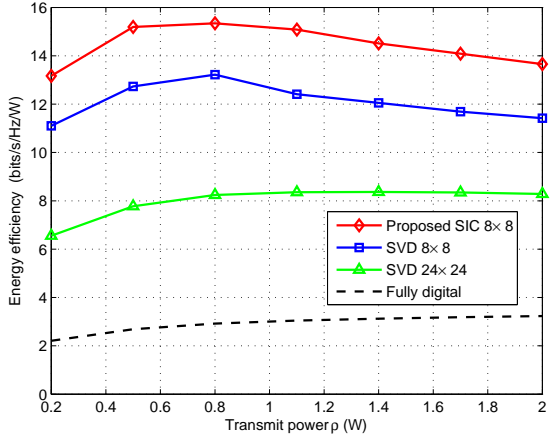}}
	\vspace{-3mm}
	\caption{EE against the transmit power. The proposed SIC-based method outperforms the traditional SVD-based methods and the fully digital method.}
	\label{Fig_EE_Pt}
\end{figure}
In Fig. \ref{Fig_EE_Pt}, we present the EE against the transmit power $\rho$. We adopt the same simulation parameters as in Fig. \ref{Fig_SE_SNR}. The noise power $\sigma^2$ is set as $0.01$. We observe that the proposed SIC-based method has higher EE than that of both the traditional SVD-based methods and the fully digital method.

\begin{figure}[tb!]
	\vspace{-0mm}
	\center{\includegraphics[width=0.95\columnwidth]{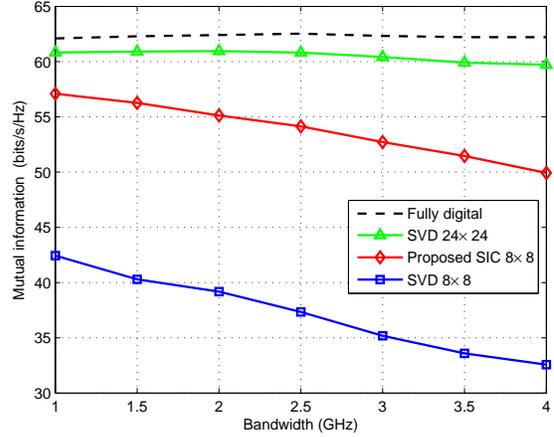}}
	\vspace{-3mm}
	\caption{MI against the bandwidth. Both the proposed ``SIC $8 \times 8$'' method and the traditional ``SVD $8 \times 8$'' method have reduced MI with a large bandwidth due to the effect of beam squint.}
	\label{Fig_SE_fs}
\end{figure}
\begin{figure}[tb!]
	\vspace{-0mm}
	\center{\includegraphics[width=0.95\columnwidth]{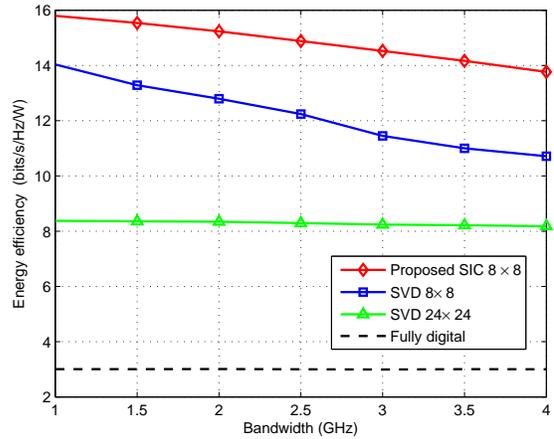}}
	\vspace{-3mm}
	\caption{EE against the bandwidth.}
	\label{Fig_EE_fs}
\end{figure}
In Fig. \ref{Fig_SE_fs} and Fig. \ref{Fig_EE_fs}, we present the MI and the EE against the bandwidth. We adopt the same simulation parameters as in Fig. \ref{Fig_SE_SNR}, except for the bandwidth. The bandwidth is set to $1\sim 4$ GHz, while the SNR is set to 20 dB. The number of selected beams is set to $N_{\rm t}^{\rm B}=N_{\rm r}^{\rm B}=24$. We observe that both the proposed ``SIC $8 \times 8$'' method and the traditional ``SVD $8 \times 8$'' method have reduced MI and EE, when the bandwidth increases. This is because the effect of beam squint becomes serious, when the bandwidth increases. This means that the leaked energy of wideband channels increases with the bandwidth.
%We also observe that the MI of the traditional ``SVD $24 \times 24$'' method remains constant. This is because 24 RF chains have captured most of the output energy of the wideband mmWave MIMO channels.

\begin{figure}[tb!]
	\vspace{-0mm}
	\center{\includegraphics[width=0.95\columnwidth]{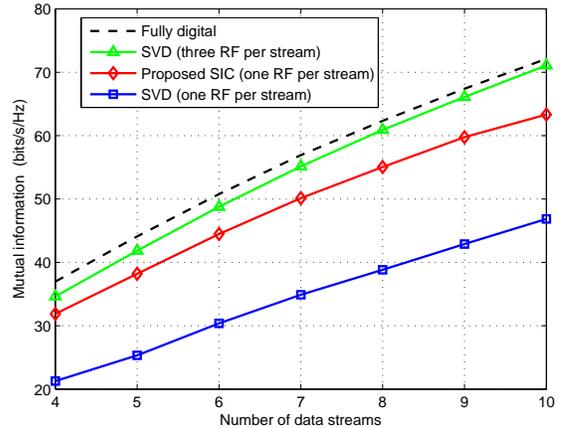}}
	\vspace{-3mm}
	\caption{MI against the number of data streams.}
	\label{Fig_SE_N_RF}
\end{figure}
\begin{figure}[tb!]
	\vspace{-0mm}
	\center{\includegraphics[width=0.95\columnwidth]{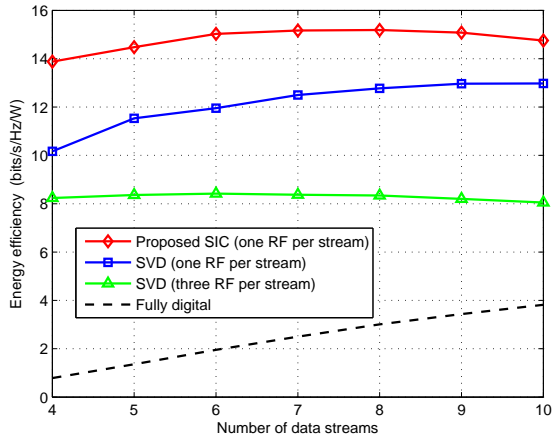}}
	\vspace{-3mm}
	\caption{EE against the number of data streams.}
	\label{Fig_EE_N_RF}
\end{figure}
In Fig. \ref{Fig_SE_N_RF} and Fig. \ref{Fig_EE_N_RF}, we present the MI and the EE against the number of data streams when SNR is set to 20 dB. We adopt the same simulation parameters as in Fig. \ref{Fig_SE_SNR} except for the number of data streams and RF chains. The number of data streams is set to $4\sim 10$. The number of RF chains is $N_{\rm t}^{\rm RF}=N_{\rm r}^{\rm RF}=N_{\rm s}$ for the proposed SIC-based method and for the traditional SVD-based method denoted by ``SVD (one RF per stream)''. We denote the traditional SVD-based method using $3N_{\rm s}$ RF chains at the transmitter and receiver as ``SVD (three RF per stream)''. We also characterize the fully digital precoding method using $64$ RF chains. Fig. \ref{Fig_SE_N_RF} shows that both the proposed SIC-based method, the traditional SVD-based methods, and fully digital method have increased MI with a large number of data streams. Furthermore, the proposed SIC-based method has higher MI than the traditional SVD-based method when the same number of RF chains is considered.
The proposed SIC-based method has higher EE than the traditional SVD-based methods and the fully digital method.

\begin{figure}[tb!]
	\vspace{-0mm}
	\center{\includegraphics[width=0.95\columnwidth]{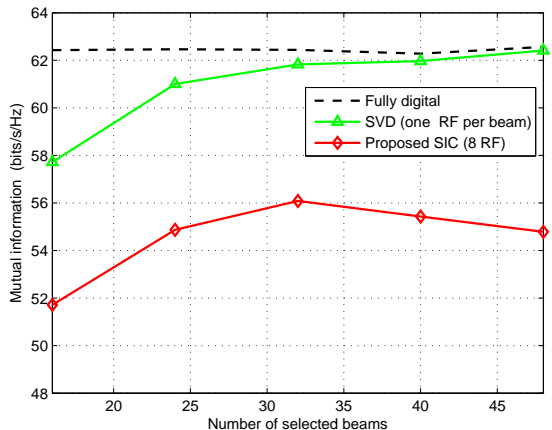}}
	\vspace{-3mm}
	\caption{MI against the number of selected beams.}
	\label{Fig_SE_NtB}
\end{figure}
\begin{figure}[tb!]
	\vspace{-0mm}
	\center{\includegraphics[width=0.95\columnwidth]{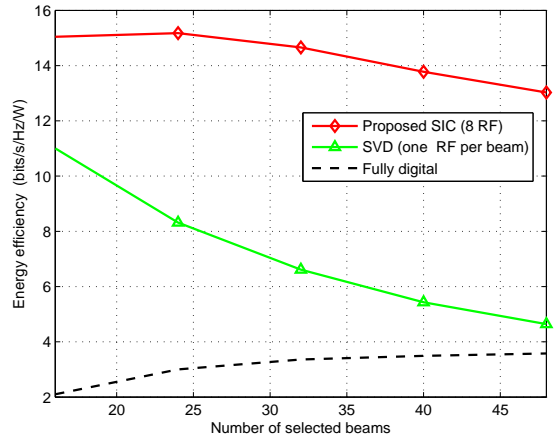}}
	\vspace{-3mm}
	\caption{EE against the number of selected beams. $N_{\rm t}^{\rm B}=N_{\rm r}^{\rm B}=24$ achieves a good trade-off between the MI performance and the hardware cost/power consumption, which is consistent with our analytical results.}
	\label{Fig_EE_NtB}
\end{figure}
In Fig. \ref{Fig_SE_NtB} and Fig. \ref{Fig_EE_NtB}, we present the MI and EE against the number of selected beams when the SNR is set to 20 dB. We adopt the same simulation parameters as in Fig. \ref{Fig_SE_SNR} except for the number of selected beams and RF chains. The number of selected beams is set to $16\sim 48$. The number of RF chains is 8 for the proposed SIC-based method and $16\sim 48$ for the traditional SVD-based method. We observe that the MI performance of the proposed SIC-based method becomes worse when the number of selected beams exceeds 32. This is because of the sub-optimality of the proposed beam selection and beamspace precoding methods, when the number of selected beams becomes very large. More explicitly, this is caused by the low-complexity sub-optimal energy-max beam selection design and the decoupling design of $\mathbf{F}_{\rm PS}$ and $\mathbf{F}_{\rm BB }$. Fig. \ref{Fig_EE_NtB} shows that $N_{\rm t}^{\rm B}=N_{\rm r}^{\rm B}=24$ strikes an attractive trade-off between the MI performance and the hardware cost/power consumption. Furthermore, we observe that the EE of the proposed SIC method is higher than that of both the traditional SVD-based method and of the fully digital method.

\section{Conclusions}\label{S5}
In this paper, we proposed a phase shifter-aided selection network for wideband mmWave MIMO systems relying on lens antenna arrays for coping with the beam squint, where each RF chain can support multiple focused-energy beams through a sub-array connected phase shifter network. We also proposed an associated TPC design, which is composed of the SIC-based beamspace precoding scheme and the low-complexity energy-max beam selection scheme. Our simulations verified that the proposed beamspace precoding and beam selection design achieves higher sum-rate and EE than its conventional counterparts. For our future research, the proposed schemes may be extended to multi-user scenarios.

\section*{Appendix}
To capture most of the channel's output energy, the number of beams selected at the transmitter and receiver can be set as
\begin{align}\label{eq_NtB1}
N_{\rm{t}}^{\rm{B}} = \left\lceil {\frac{L{{N_{\rm{t}}}B}}{{2{f_{\rm c}}}}} \right\rceil,
N_{\rm{r}}^{\rm{B}} = \left\lceil {\frac{L{{N_{\rm{r}}}B}}{{2{f_{\rm c}}}}} \right\rceil.
\end{align}
\begin{IEEEproof}
We first consider the case of single-path channels, i.e., $ L = 1$. Then we will expand the results to the case of multi-path channels having $ L \ge 2$.
For the single-path channels, the beamspace channel ${{\bf{H}}_{\rm{b}}}\left[ k \right]$ can be simplified as
\begin{align}
{{\bf{H}}_{\rm{b}}}\left[ k \right] = \beta {{\bf{\bar a}}_{\rm{r}}}\left( {{\phi _{\rm{r}}^k}} \right){\bf{\bar a}}_{\rm{t}}^{\rm H}\left( {{\phi _{\rm{t}}^k}} \right),
\end{align}
where $\beta$ is the path gain, $\phi _{\rm{t}}^k$ and $\phi _{\rm{r}}^k$ are the spatial AoD and AoA.
The antenna array response vectors $ {{\bf{\bar a}}_{\rm{t}}}\left( {{\phi _{\rm{t}}^k}} \right)$ and ${\bf{\bar a}}_{\rm{r}}\left( {{\phi _{\rm{r}}^k}} \right)$ are given by
\begin{align}
 &{{\bf{\bar a}}_{\rm{t}}}\left( {{\phi _{\rm{t}}^k}} \right) =\left[{\Xi_{{N_{\rm{t}}}}}\left({{\phi _{\rm{t}}^k}}-\bar\phi_{\rm{t}}^1 \right),\cdots,{\Xi_{{N_{\rm{t}}}}}\left({{\phi _{\rm{t}}^k}}-\bar\phi_{\rm{t}}^{N_{\rm t}} \right) \right]^{\rm T},\\
& {{\bf{\bar a}}_{\rm{r}}}\left( {{\phi _{\rm{r}}^k}} \right) =\left[{\Xi_{{N_{\rm{r}}}}}\left({{\phi _{\rm{r}}^k}}-\bar\phi_{\rm{r}}^1 \right),\cdots,{\Xi_{{N_{\rm{r}}}}}\left({{\phi _{\rm{r}}^k}}-\bar\phi_{\rm{r}}^{N_{\rm r}} \right) \right]^{\rm T}.
\end{align}
The selection matrix at the transmitter is obtained by choosing the largest $N_{\rm t}^{\rm B}$ diagonal element of $\frac{1}{K}\sum\limits_{k = 1}^K {{\bf{H}}_{\rm{b}}^{\rm{H}}\left[ k \right]}{{\bf{H}}_{\rm{b}}}\left[ k \right] $, which is given by
\begin{align}\label{eq_diag}
\frac{{{\beta ^2}}}{K}\sum\limits_{k = 1}^K {\left[ {\begin{array}{*{20}{c}}
		{\Xi _{{N_{\rm{t}}}}^2\left( {\phi _{\rm{t}}^k - \bar \phi _{\rm{t}}^1} \right)}&{}&{}\\
		{}& \ddots &{}\\
		{}&{}&{\Xi _{{N_{\rm{t}}}}^2\left( {\phi _{\rm{t}}^k - \bar \phi _{\rm{t}}^{{N_{\rm{t}}}}} \right)}
		\end{array}} \right]} .
\end{align}

\newcounter{mytempeqncnt}
\textbf{\begin{figure*}[!t]
		% ensure that we have normalsize text
		\normalsize
		% Store the current equation number.
		\setcounter{mytempeqncnt}{\value{equation}}
		% Set the equation number to one less than the one
		% desired for the first equation here.
		% The value here will have to changed if equations
		% are added or removed prior to the place these
		% equations are referenced in the main text.
		\setcounter{equation}{44}
		\begin{equation}
		{\rm{diag}}\left( {\frac{1}{K}\sum\limits_{k = 1}^K {{\bf{H}}_{\rm{b}}^{\rm{H}}\left[ k \right]} {{\bf{H}}_{\rm{b}}}\left[ k \right]} \right) = \frac{1}{K}\sum\limits_{k = 1}^K {\left[ {\begin{array}{*{20}{c}}
				{\sum\limits_{\ell  = 1}^L {\sum\limits_{{s_\ell } = 1}^{{S_\ell }} {{\beta _{{s_{\ell ,k}}}}\Xi _{{N_{\rm{t}}}}^2\left( {\phi _{\rm{t}}^{{s_\ell },k} - \bar \phi _{\rm{t}}^1} \right)} } }&{}&{}\\
				{}& \ddots &{}\\
				{}&{}&{\sum\limits_{\ell  = 1}^L {\sum\limits_{{s_\ell } = 1}^{{S_\ell }} {{\beta _{{s_{\ell ,k}}}}\Xi _{{N_{\rm{t}}}}^2\left( {\phi _{\rm{t}}^{{s_\ell },k} - \bar \phi _{\rm{t}}^{{N_{\rm{t}}}}} \right)} } }
				\end{array}} \right]} .
		\end{equation}
		% Restore the current equation number.
		\setcounter{equation}{\value{mytempeqncnt}}
		% IEEE uses as a separator
		\hrulefill
		% The spacer can be tweaked to stop underfull vboxes.
		\vspace*{4pt}
\end{figure*}}

Observe that ${\Xi _N}^2\left( x \right)$ has the following characteristics: $ {\Xi _N}\left( x \right)^2 \approx 0$ when $|x|\gg 1/N$. Thus the diagonal element $\Xi _{{N_{\rm{t}}}}^2\left( {{\phi _{\rm{t}}^k} - \bar \phi _{\rm{t}}^n} \right)$ is a non-negligible value only when ${\phi _{\rm{t}}^k} \approx \bar \phi _{\rm{t}}^n$. Without loss of generality, we assume that ${\phi _{\rm{t}}^{K/2}} = \bar \phi _{\rm{t}}^{n^\star}>0 $, where ${\phi _{\rm{t}}^{K/2}}$ is the spatial AoD at the central subcarrier, which can be expressed as ${\phi _{\rm{t}}^{K/2}} = \left( {d{f_{\rm{c}}}/c} \right)\sin {\theta _{\rm{t}}}$, where $\theta _{\rm{t}}$ is the physical path AoD. Note that ${f_k} ={f_{\rm c}}\left( {1 + \frac{B}{{{f_{\rm c}}K}}\left( {k - 1 - \frac{{K - 1}}{2}} \right)} \right)$ for $k=1,2,\cdots, K$. Therefore, the spatial AoD ${\phi _{\rm{t}}^k}$ at the $k$-th subcarrier can be bounded as
\begin{align}\label{eq_phit}
{\phi _{\rm{t}}^{K/2}}\left( {1 - \alpha } \right) \le {\phi _{\rm{t}}^k} \le {\phi _{\rm{t}}^{K/2}}\left( {1 + \alpha } \right),
\end{align}
where $\alpha  =\frac{{B\left( {K - 1} \right)}}{{2K{f_{\rm{c}}}}}$. If ${\phi _{\rm{t}}^k}=\bar \phi _{\rm{t}}^n=\frac{1}{{{N_{\rm{t}}}}}\left( {n - \frac{{{N_{\rm{t}}} + 1}}{2}} \right)$,
(\ref{eq_phit}) is rewritten as,
\begin{align}\label{eq_ineq1}
\!\!\frac{{\left( {{n^ \star }\!\! - \!\frac{{{N_{\rm{t}}} + 1}}{2}} \right)\left( {1 \!-\! \alpha } \right)}}{{{N_{\rm{t}}}}}\! \le\! \frac{{{n\! - \!\frac{{{N_{\rm{t}}} + 1}}{2}}}}{{{N_{\rm{t}}}}}\! \le \!\frac{{\left( {{n^ \star }\! \!-\! \frac{{{N_{\rm{t}}} + 1}}{2}} \right)\left( 1 \! + \!\alpha \right)}}{{{N_{\rm{t}}}}},
\end{align}
which can be simplified as
\begin{align}\label{eq_ineq2}
{n^ \star }\! -\! \alpha \left( {{n^ \star }\! -\! \frac{{{N_{\rm{t}}} + 1}}{2}} \right) \le n \le {n^ \star }\! + \!\alpha \left( {{n^ \star }\! -\! \frac{{{N_{\rm{t}}} + 1}}{2}} \right).
\end{align}
This implies that ${\phi _{\rm{t}}^k}=\bar \phi _{\rm{t}}^n$ may happen when $n$ satisfies (\ref{eq_ineq2}). Hence, the number of non-negligible diagonal elements of $\frac{1}{K}\sum\limits_{k = 1}^K {{\bf{H}}_{\rm{b}}^{\rm{H}}\left[ k \right]}{{\bf{H}}_{\rm{b}}}\left[ k \right] $ is smaller than $2\alpha \left( {{n^ \star } - \frac{{{N_{\rm{t}}} + 1}}{2}} \right)$. Since $\frac{{{N_{\rm{t}}} + 1}}{2} \le {n^ \star } \le {N_{\rm{t}}}$, the number of non-negligible diagonal elements is smaller than $\alpha \left( {{N_{\rm{t}}}{\rm{ - }}1} \right){\rm{ = }}\frac{{B\left( {K - 1} \right)\left( {{N_{\rm{t}}}{\rm{ - }}1} \right)}}{{2K{f_{\rm{c}}}}}$. Therefore, the number of beams  $N_{\rm t}^{\rm B}$ selected at the transmitter can be approximated by (\ref{eq_NtB1}).

For multi-path channels, the beamspace channel $\mathbf{H}_{\rm b}[k]$ is modeled as
\begin{align}\label{eq_Hbk3}
\mathbf{H}_{\rm b}[k]=\sum_{\ell=1}^{L}\sum_{s_{\ell}=1}^{S_{\ell}}\beta_{s_{\ell,k}}
\bar{\mathbf{a}}_{\rm r}\left( \phi_{\rm r}^{s_{\ell},k}\right)
\bar{\mathbf{a}}_{\rm t}^{\rm H}\left( {\phi _{\rm{t}}^{{s_\ell,k}}} \right).
\end{align}
The expression of the diagonal elements of $\frac{1}{K}\sum\limits_{k = 1}^K {{\bf{H}}_{\rm{b}}^{\rm{H}}\left[ k \right]}{{\bf{H}}_{\rm{b}}}\left[ k \right] $ in (\ref{eq_diag}) becomes very complicated. Observe that the more non-zero elements ${{\bf{H}}_{\rm{b}}^{\rm{H}}\left[ k \right]}$ has, the more non-negligible diagonal elements $\frac{1}{K}\sum\limits_{k = 1}^K {{\bf{H}}_{\rm{b}}^{\rm{H}}\left[ k \right]}{{\bf{H}}_{\rm{b}}}\left[ k \right] $ involves. Thus, we consider the worst case, where the AoD spread of each scatterer is small and non-overlapped. This means that the beamspace channel contains the highest number of non-zero elements. In this case, the array response vectors of the subpaths in the different scatterers are asymptotically orthogonal and the array response vectors of subpaths in the same scatterers are similar. Then (\ref{eq_diag}) can be rewritten for multi-path channels as (45). Similar to the case of single-path channels, the number of non-negligible diagonal elements is smaller than $L\alpha \left( {{N_{\rm{t}}}{\rm{ - }}1} \right){\rm{ = }}\frac{{LB\left( {K - 1} \right)\left( {{N_{\rm{t}}}{\rm{ - }}1} \right)}}{{2K{f_{\rm{c}}}}}$.
Therefore, the number of beams $N_{\rm t}^{\rm B}$ selected at the transmitter can be approximated by (\ref{eq_NtB1}).

For the receiver side, the proof is similar to that at the transmitter, thus we omit it due to the space limitations.
\end{IEEEproof}
%\section*{Appendix II}
	
%	\newpage
	\bibliographystyle{IEEEtran}
	\bibliography{IEEEabrv,Refference}

\begin{IEEEbiography}[{\includegraphics[width=1in,height=1.25in,clip,keepaspectratio]{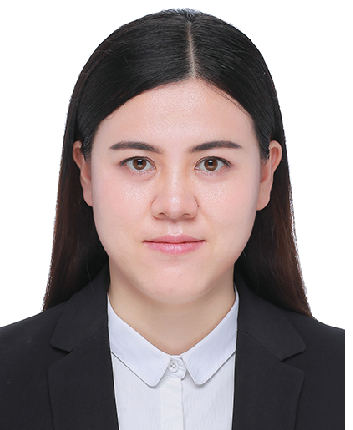}}]
		{Wenqian Shen} received the B.S. degree from Xi'an Jiaotong University, Shaanxi, China in 2013 and the Ph.D. degree from Tsinghua University, Beijing, China. She is currently a Post-Doctoral Research Fellow with the School of Information and Electronics, Beijing Institute of Technology, Beijing, China. Her research interests include massive MIMO and mmWave/THz communications. She has published several journal and conference papers in IEEE Transaction on Signal Processing, IEEE Transaction on Communications, IEEE Transaction on Vehicular Technology, IEEE ICC, etc. She has won the IEEE Best Paper Award at the IEEE ICC 2017.
\end{IEEEbiography}
	
\begin{IEEEbiography}[{\includegraphics[width=1in,height=1.25in,clip,keepaspectratio]{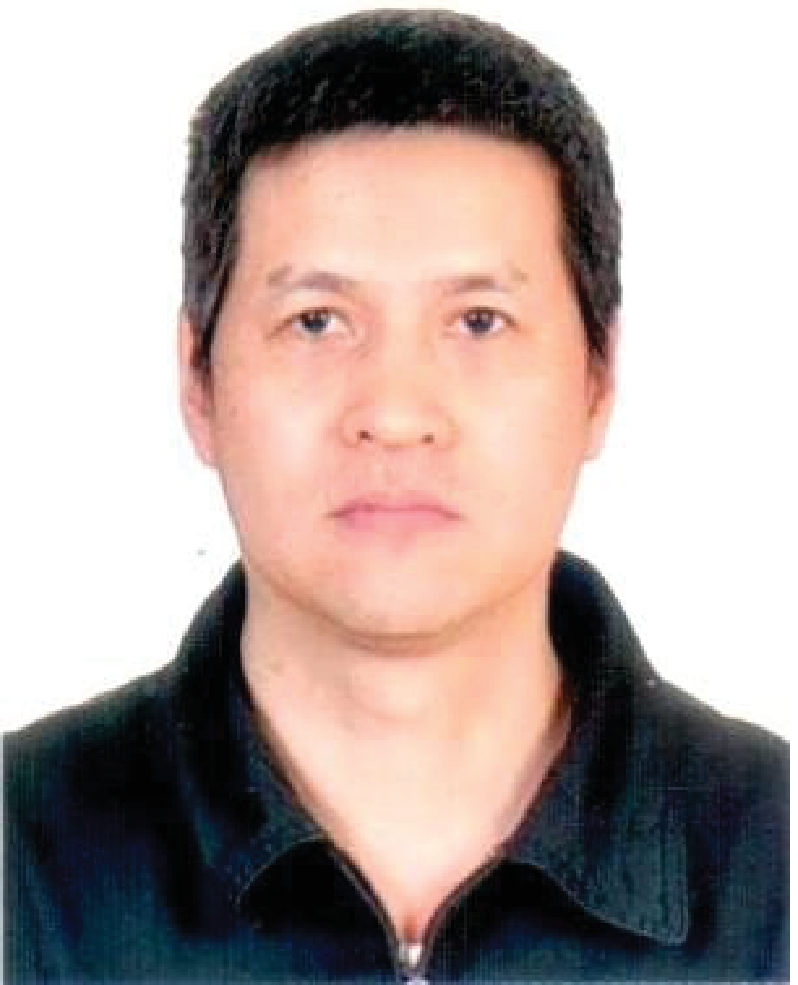}}]
		{Xiangyuan Bu} received the B.E. and Ph.D. degree in communications engineering from the Beijing Institute of Technology (BIT), Beijing, China, in 1987 and 2007 respectively. He is currently a professor in the School of Information and Electronics, BIT. His current research interests include digital signal processing, channel coding theory, MIMO system, space time signal processing and satellite communications.
\end{IEEEbiography}

\begin{IEEEbiography}[{\includegraphics[width=1in,height=1.25in,clip,keepaspectratio]{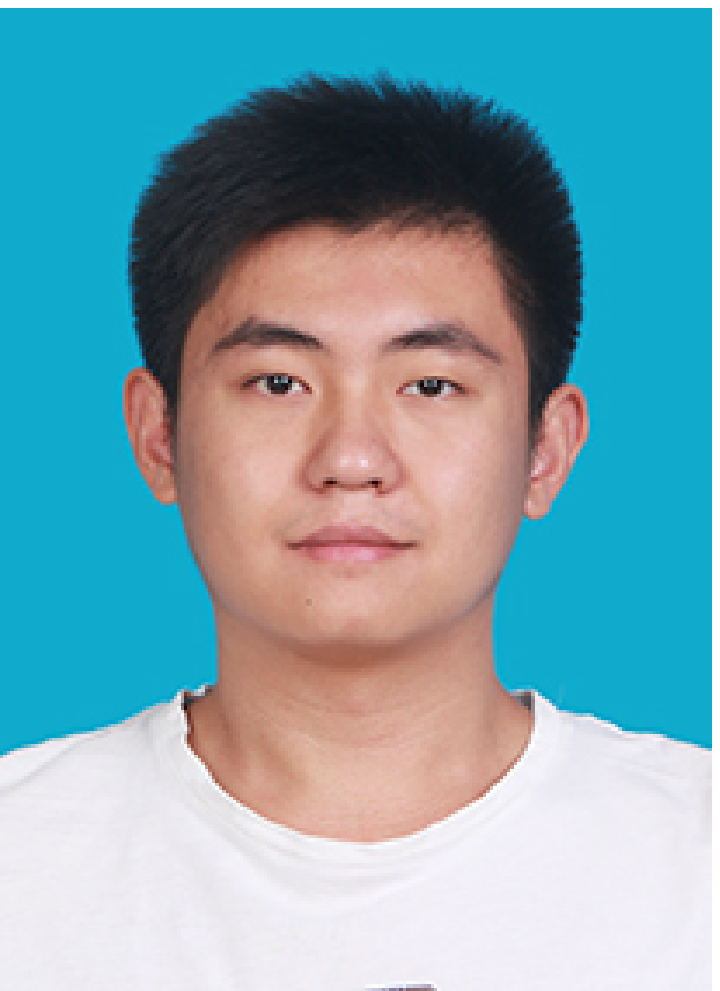}}]
    	{Xinyu Gao}(S'14) received the B.E. degree of Communication	Engineering from Harbin Institute of Technology, Heilongjiang, China in 2014 and the PhD degree of Electronic Engineering from Tsinghua University, Beijing, China in 2019 (with the highest honor). He is currently working as a senior engineer for Huawei Technology, Beijing, China. His research interests include massive MIMO and mmWave communications, with the emphasis on signal processing. He has published more than 20 IEEE journal and conference papers, such as IEEE Journal on Selected Areas in Communications, IEEE Transaction on Signal Processing, IEEE ICC, IEEE GLOBECOM, etc. He has won the WCSP Best Paper Award and the IEEE ICC Best Paper Award in 2016 and 2018, respectively.
\end{IEEEbiography}

\begin{IEEEbiography}[{\includegraphics[width=1in,height=1.25in,clip,keepaspectratio]{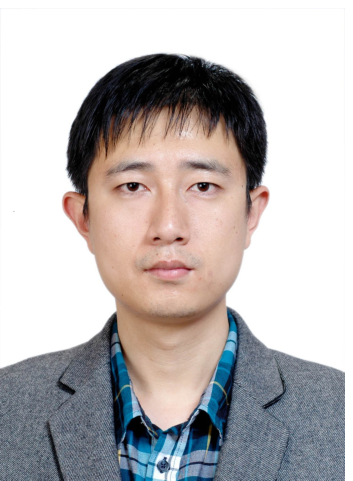}}]
		{Chengwen Xing} (S'08-M'10) received his B.Eng. degree from Xidian University, Xian, China, in 2005,
		and his Ph.D. degree from the University of Hong Kong, Hong Kong,  China, in 2010. 	Since September 2010, he has been with the School of Information and Electronics, Beijing Institute of Technology, Beijing, China, where he is currently a Full Professor. From September 2012 to December 2012, he was a visiting scholar at the University of Macau. His current research interests include statistical signal processing, convex optimization, multivariate statistics, combinatorial optimization, massive MIMO systems, and high frequency band communication systems. Prof. Xing is an Associate Editor for the IEEE Transactions on Vehicular Technology, KSII Transactions on Internet and Information Systems, Transactions on Emerging Telecommunications Technologies, and China Communications.
\end{IEEEbiography}
	
\begin{IEEEbiography}[{\includegraphics[width=1in,height=1.25in,clip,keepaspectratio]{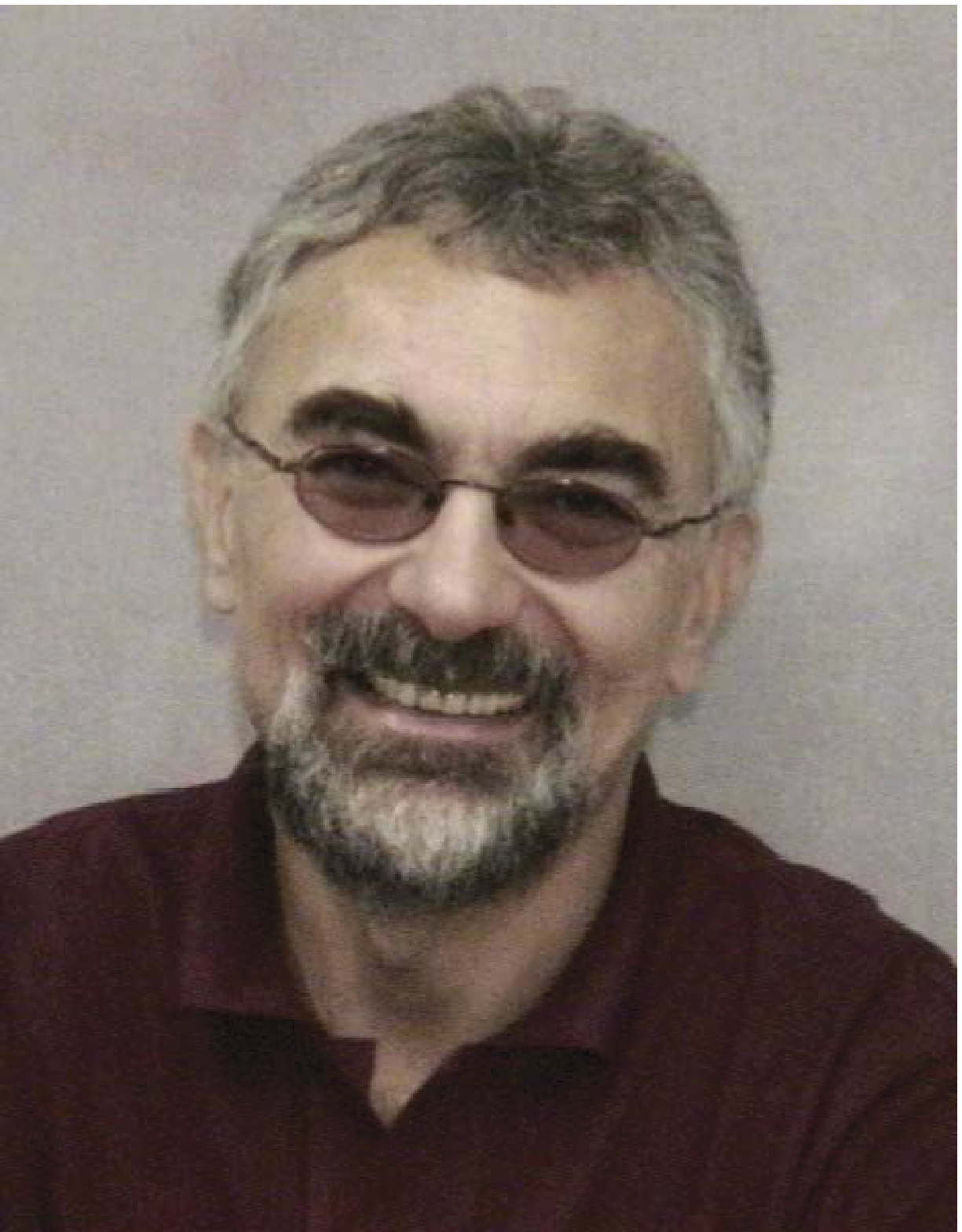}}]
		{Lajos Hanzo} (FREng, F'04, FIET, Fellow of EURASIP) received his 5-year degree in Electronics in 1976 and his doctorate in 1983 from the Technical University of Budapest.  In 2009 he was awarded an honorary doctorate by the Technical University of Budapest and in 2015 by the University of Edinburgh.  In 2016 he was admitted to the Hungarian Academy of Science. During his 40-year career in telecommunications he has held various research and academic posts in Hungary, Germany and the UK. Since 1986 he has been with the School of Electronics and Computer Science, University of Southampton, UK, where he holds the chair in telecommunications.  He has successfully supervised 119 PhD students, co-authored 18 John Wiley/IEEE Press books on mobile radio communications totalling in excess of 10,000 pages, published 1800+ research contributions at IEEE Xplore, acted both as TPC and General Chair of IEEE conferences, presented keynote lectures and has been awarded a number of distinctions. Currently he is directing a	60-strong academic research team, working on a range of research projects in the field of wireless multimedia communications sponsored by industry, the Engineering and Physical Sciences Research Council (EPSRC) UK, the European Research Council's Advanced Fellow Grant and the Royal Society's Wolfson Research Merit Award.  He is an enthusiastic supporter of industrial and academic liaison and he offers a range of industrial courses.  He is also a Governor of the IEEE ComSoc and VTS.  He is a former Editor-in-Chief of the IEEE Press and a former Chaired Professor also at Tsinghua University, Beijing.  For further information on research in progress and associated publications please refer to http://www-mobile.ecs.soton.ac.uk
\end{IEEEbiography}

\end{document}